\documentclass{IEEEtran}
\IEEEoverridecommandlockouts
\usepackage{cite}
\usepackage{algorithmic}
\usepackage{xcolor}
\usepackage{enumerate}
\usepackage{amsfonts,amsmath,amssymb,amsfonts}
\usepackage{amsthm}
\usepackage{algorithmic}
\usepackage{optidef}
\usepackage{algorithm}
\usepackage{mathrsfs}
\usepackage{graphicx}
\usepackage{textcomp}
\usepackage{subfigure}
\usepackage[squaren]{SIunits}
\usepackage[section]{placeins}

\def\BibTeX{{\rm B\kern-.05em{\sc i\kern-.025em b}\kern-.08em
    T\kern-.1667em\lower.7ex\hbox{E}\kern-.125emX}}
\begin{document}
\newtheorem{theorem}{\textbf{Theorem}}
\newtheorem{lemma}{\textbf{Lemma}}
\bibliographystyle{IEEEtran}

\title{Computational Offloading in Semantic-Aware Cloud-Edge-End Collaborative Networks}
\author{\IEEEauthorblockN{Zelin Ji,~\IEEEmembership{Graduate Student Member,~IEEE} and
Zhijin Qin,~\IEEEmembership{Senior Member,~IEEE}}
\thanks{Zelin~Ji is with School of Electronic Engineering and Computer Science, Queen Mary University of London, London E1 4NS, U.K. (email: z.ji@qmul.ac.uk).}
\thanks{Zhijin~Qin is with Department of Electronic Engineering, Tsinghua University, Beijing, China. She is also with the Beijing National Research Center for Information Science and Technology, Beijing, China, and the State Key Laboratory of Space Network and Communications, Beijing, China. (email: qinzhijin@tsinghua.edu.cn).
}
}
\maketitle

\begin{abstract}
The trend of massive connectivity pushes forward the explosive growth of end devices. The emergence of various applications has prompted a demand for pervasive connectivity and more efficient computing paradigms. On the other hand, the lack of computational capacity of the end devices restricts the implementation of the intelligent applications, and becomes a bottleneck of the multiple access for supporting massive connectivity. Mobile cloud computing (MCC) and mobile edge computing (MEC) techniques enable end devices to offload local computation-intensive tasks to servers by networks. In this paper, we consider the cloud-edge-end collaborative networks to utilize distributed computing resources. Furthermore, we apply task-oriented semantic communications to tackle the fast-varying channel between the end devices and MEC servers and reduce the communication cost. To minimize long-term energy consumption on constraints queue stability and computational delay, a Lyapunov-guided deep reinforcement learning hybrid (DRLH) framework is proposed to solve the mixed integer non-linear programming (MINLP) problem. The long-term energy consumption minimization problem is transformed into the deterministic problem in each time frame. The DRLH framework integrates a model-free deep reinforcement learning algorithm with a model-based mathematical optimization algorithm to mitigate computational complexity and leverage the scenario information, so that improving the convergence performance. Numerical results demonstrate that the proposed DRLH framework achieves near-optimal performance on energy consumption while stabilizing all queues.

\end{abstract}


\section{Introduction} 

 

In the scope of next-generation wireless networks with computation functions, the emerging applications and techniques put forward the demands on massive connectivity and computational capacity~\cite{9615110}. The next-generation wireless networks should efficiently and reliably facilitate massive device connectivity while guaranteeing the quality of service (QoS). The massive connectivity techniques include the transmission techniques that support the enormous number of user devices, the communication resource allocation techniques that coordinate the physical resource and perform the power control function, and the computational resource scheduling that enhances the ability to integrate artificial intelligent models and realize the intelligent networks~\cite{chen2023signal}. 

However, the massive connectivity demand for 6G and beyond communications is far from being satisfied, which calls for a new information transmission paradigm to be developed. Semantic communications have been attracting more and more attention with the potential to enhance communications by reducing the transmission overhead and exchanging information more efficiently. Different from the source encoding in conventional transmissions, semantic communications extract the semantic information and encode the information to semantic features, which are transmitted by the physical channels~\cite{qin2022semantic, 9830752}. Meanwhile, the challenge to realizing massive connectivity in semantic-aware networks is the limited computational capability of battery-powered mobile devices, which inspires us to leverage the rich computational resources at the servers and collaborate the resources among the whole wireless networks, to enhance the multiple access and realize the massive connectivity for the large-scale networks. 

Mobile cloud computing (MCC) has emerged as the solution for end users to offload computations and task workloads to compute-intensive cloud centers. Nevertheless, the cloud center faces challenges raised by the centralized service mode, e.g., the restricted bandwidth and network congestion, etc~\cite{cloud_latency}. To tackle the aforementioned challenges associated with MCC, the paradigm of Mobile Edge Computing (MEC) has emerged as a novel computational approach by enabling end users to connect to edge servers instead of cloud servers, thereby substantially diminishing the communication costs incurred in task offloading due to the proximity of these edge servers to end devices. In~\cite{10102429}, Hoang~\emph{et al.} parallel computation between end devices and cloud and edge servers. Nevertheless, the computational capability of MEC servers is inferior to that of MCC servers. Additionally, MCC servers are typically positioned at centralized base stations (BS), facilitating swift access to channel state information (CSI) for effective management of fast-varying channels. On the other hand, MEC servers lack the capability for real-time CSI feedback, so a more robust transmission technique with a stable performance guarantee needs to be applied~\cite{9099242}.

\subsection{Cloud-Edge-End Collaboration Intelligence}

The limited resources in cloud-edge-end networks have become a bottleneck for the multiple access of edge computing~\cite{8016573}. To improve the communication performance and computational capability, various studies have focused on resource allocation techniques. These techniques optimize the user scheduling~\cite{8851249}, bandwidth allocation~\cite{9194337}, transmit power~\cite{9210812}, and joint optimization~\cite{10415196} for cloud-edge-end networks. In~\cite{9556549}, Zhou~\emph{et al.} consider both computation and communication resources for ESs and local devices, jointly optimizing the transmit power and computational capability for local devices, and achieving the energy-efficient resource allocation. Kai~\emph{et al.}~\cite{9171865} and Ding~\emph{et al.}~\cite{9954169} applied the successive convex approximation approach for the optimization of task and computation offloading performance. Nevertheless, these works rely on stable channel state information (CSI), which is impractical for advanced communication systems.

To overcome the challenge of the fast-varying channels and to make decisions under unstable CSI, Min~\emph{et al.}~\cite{8598893} proposed an offloading scheme relying on reinforcement learning (RL), so that the IoT device can make decisions without the foreknowledge of the servers. Meanwhile, the presence of fast-varying channels and limited bandwidth resources necessitates communication techniques that are robust to fast channel variations while reducing data traffic for computational offloading.

The deployment and applications of artificial intelligence (AI) and large language model (LLM) based tasks put higher requirements on the computational resources. Nevertheless, due to the lack of essential computational capability and power supply of end devices, executing intelligent high-level tasks on end devices is a challenge. The cloud-edge-end network provides a promising solution, which enables the end users to offload local task information to cloud or edge servers and leverage the rich computational resources. 

Generally speaking, most intelligence tasks are based on the deep neural network (DNN), and the large number of weight connections led to enormous floating point operations. To address the computational complexity and caching issue, Chen~\emph{et al.}~\cite{9862981} joint optimize the task software caching update and computation offloading. Offloading part of the DNN calculations to servers, i.e., DNN partitioning, provided a method to transfer the computation pressure of the end devices to servers. Teerapittayanon~\emph{et al.} proposed a novel model to accelerate the inference by classifying samples from early layers and designing the early exit scheme. By combining the early exit scheme and edge computing, Zhang~\cite{3397315} proposed a method to find the optimal partition of the DNN to minimize the latency of the inference process. Furthermore, Wang~\emph{et al.}~\cite{9685056} proposed an attention-based learning approach to extract the semantic features of the transmitted text and allocate the resource blocks to optimize the semantic similarity. Nevertheless, the current DNN partition works assume that the DNN layers can be divided arbitrarily, and the parameters of the DNN layers can be transmitted and received with no error by the conventional transmission techniques, which is unpractical when facing fast-varying wireless channel situations. To address this challenge, a scheme that jointly considers the DNN partitioning and transmission between the end devices and servers needs to be investigated and implemented.

\subsection{Semantic-Aware Computational Offloading Systems}

One way to realize the DNN partitioning with communications is semantic communications, which is categorized into the second level of communication. Instead of transmitting the source raw data by conventional approach with source coding and channel coding, the DNN-based semantic communications extract the semantic information of the raw source data and only transmit the essential "meaning" or task-related information for the receivers to reconstruct the data, which enhances the robustness and reduce the overheads of the communication, and further cope with the computation capacity and fast varying channel challenges~\cite{10183794, 9398576}.

Particularly, we adopt the DeepSC communication model proposed by Xie~\emph{et al.}~\cite{9252948}, where the pre-trained semantic encoder and decoder are applied to extract and reconstruct the semantic information and implement the semantic communications. Yan~\emph{et al.} designed new metrics for semantic communications, including the semantic rate, semantic spectrum efficiency, and semantic quality of experience (QoE). Specifically, the semantic spectrum efficiency was maximized for text-based semantic communication systems in~\cite{9763856}, and the QoE was optimized for multi-modal semantic communications in~\cite{10001594}. To enhance the computational capacity of the end device, Ji~\emph{et al.}~\cite{ji2024resource} proposed a semantic-aware task offloading system, where the semantic decoders are implemented at the ES, relieving the computational burden at the end devices. However, this paper only considers short-term energy consumption and does not consider the long-term influence of the task queue. Moreover, this paper only considers a single edge server and only optimizes the energy consumption of the local users. Nevertheless, it is essential to consider the resource costs by other resource entities, and jointly collaborate the computation resources of the entire cloud-edge-end networks, which can be abstracted as a computing network system where the heterogeneous cloud, edge, and end computing entities are interconnected~\cite{10387520}, so that leveraging and coordinating the resources for supporting the massive connectivity.

\subsection{Unified Resource Optimization Framework}
To manage the heterogeneous computing resources and the communication resources, a unified resource optimization framework is needed. However, the offloading policy of different end devices and the tasks are strongly coupled, leading to significant complexity when the number of end devices is large. Meanwhile, the continuous resource optimization variables, e.g., the computation frequency and the transmit power, are jointly optimized with the binary offloading policy, which renders the computational offloading optimization problem a mixed-integer nonlinear programming (MINLP) problem~\cite{9442308}. 

To solve the MINLP problem, the model-based mathematical optimization algorithms and the model-free machine learning approaches can be leveraged~\cite{8901995}. The model-based optimization algorithms are based on the physical system, and apply mathematical models and tools to provide the optimal solution with a theoretical guarantee of convergence. However, such algorithms are usually computationally expensive, further affecting the limited computational resources of the end devices. Furthermore, these conventional model-based methods usually require a large number of long-term iterations until achieve a good performance, which cannot satisfy the task execution latency requirement and is impractical to the fast-varying wireless channels.

On the other hand, data-driven solutions, e.g., machine learning approaches provide a promising solution to tackle the complexity issue. Specifically, the deep reinforcement learning (DRL) algorithm enables the agents to learn the offloading policy and access the computational and communication resources by interacting with the environment. The DRL models can be applied for online real-time control, reducing the computational cost and latency significantly.

Nevertheless, the model-free algorithms may suffer from unstable performance and slow convergence~\cite{9679802}. To combine the advantage of the performance guarantee of the model-based conventional optimization problem and the low complexity of the model-free algorithms, Bi~\emph{et al.}~\cite{9449944} propose an integration structure of model-based optimization critic module that can enhance the robustness, and the model-free DRL actor module that can reduce the computational complexity. However, the above works only consider the conventional task offloading scheme to a single server, neglecting the heterogeneity of MCC and MEC servers.

{To implement the intelligent signal processing and learning, and coordinate the resources in the massive connectivity scenarios, we present a cloud-edge-end collaborative semantic-aware computational offloading network in this paper.} Specifically, the semantic-aware offloading and the conventional offloading scheme are applied to take advantage of the MEC and MCC servers. The offloading policy, offloading volume, and computing capacity are jointly optimized. To realize online optimization with low complexity, a Lyapunov-guided deep reinforcement learning based hybrid framework is proposed. To the best of our knowledge, this is the first work that performs the joint optimization of the long-term energy consumption for a cloud-edge-end collaborative semantic-aware computational offloading network. {Nonetheless, we faces the issues of the transmission techniques, the computational capacity requirement, and the low-complexity resource optimization requirement to be addressed.}

\begin{itemize}

 \item {The limited computational capacity and memory of the end devices. How to collaborate the computational resource of end devices, the MEC server, and the MCC server?}
 \item {{The massive connectivity requires the efficient transmission techniques. How to implement advanced semantic communication in the proposed collaborative networks?}}
  \item {A joint offloading scheme need to be designed to guarantee the offloading performance of the heterogeneous MCC and MEC servers. How to design the offloading policy to reduce energy consumption?}
\item {How to minimize the long-term energy consumption of the proposed network? How to jointly optimize the discrete and continuous resources? How to design a low-complexity online optimization algorithm?}

\end{itemize}

The contributions of this paper that address the above challenges are concluded as follows.
\begin{enumerate}
\item {\textbf{Cloud-edge-end collaborate offloading}: In this paper, a cloud-edge-end collaborate semantic-award computational offloading system is considered. The resources of the cloud, edge servers, and end devices are unified and jointly managed to support massive connectivity. To achieve this, a multi-tier management algorithm is proposed.}

\item{\textbf{Semantic-aware computational offloading}: To utilize the computational capacity of the end devices and improve the transmission robustness of the offloading process, the semantic communication technique is applied to extract the task-related information. The task execution can be selected from three modes, including local execution, semantic-aware offloading, and source data offloading.}

\item {\textbf{Lyapunov-guided problem transform}: 
To enhance the long-term energy efficiency of the system, we apply the Lyapunov-guided algorithm to convert the long-term optimization problem into a deterministic problem within each discrete time frame, which obviates the need for future information and can be formulated as the online optimization problem.}

\item {\textbf{Deep reinforcement learning based hybrid (DRLH) framework}: To solve the joint optimization problem with discrete and continuous variables, a hybrid algorithm is proposed to utilize the advantage of different algorithms. Specifically, to cope with the complexity issue, a deep reinforcement learning algorithm is proposed for the sub-optimal solutions. To cope with the long-term objectives of the system while guaranteeing the constraints on the task queue stability and latency, the Lyapunov optimization is applied.}



\end{enumerate}

The rest of this article is organized as follows. Section~\uppercase\expandafter{\romannumeral2} presents the offloading and computation models for different modes of task executions. In Section~\uppercase\expandafter{\romannumeral3}, the energy consumption minimization problem is formulated and the Lyapunov-guided algorithm is applied to transform the long-term optimization problem into deterministic problems. The hybrid algorithm to solve the formulated problem is presented in Section~\uppercase\expandafter{\romannumeral4}, where the detailed training process is introduced. The numerical simulation results are demonstrated in Section~\uppercase\expandafter{\romannumeral5}. The conclusion is drawn in Section~\uppercase\expandafter{\romannumeral6}.

\section{System Model}
\label{chp6:system}

To utilize the powerful computational capacity of the MCC servers and the low latency of the MEC servers, while overcoming the disadvantage of the cloud and edge servers, we consider a cloud-edge-end collaborative network with multiple users. Specifically, we consider two servers, one located at a macro base station as the MCC server, and the other as a MEC server, providing extra and proximal computing services for the end users. The end devices can access the MCC and MEC servers to offload the tasks at the same time. For tasks offloaded to the MEC server, we apply the DeepSC model to realize the semantic-aware computational offloading scheme, to relieve the MEC's computational pressure and cope with the unstable channels. Meanwhile, for the tasks offloaded to the MCC server, the conventional computational offloading technique is adopted to take advantage of the rich computational resources and lower the processing latency from the end device side.

\begin{figure}[t]
\centering
\includegraphics[width=\columnwidth]{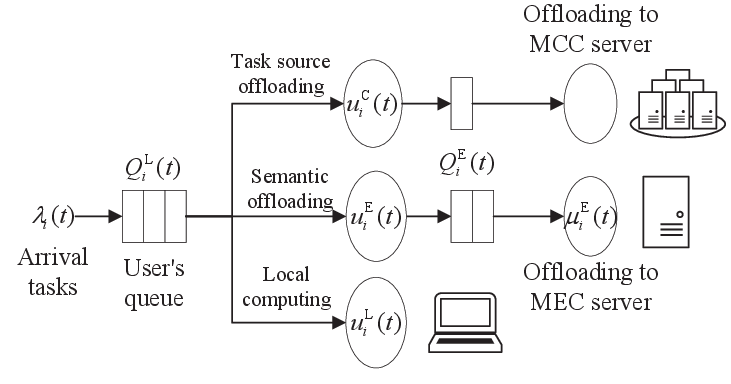}
\caption{Queuing and task execution model of the proposed cloud-edge-end network.}
\label{chp6:queue}
\end{figure}

We define the set of UEs as ${\cal U} = \{U_i|i=1,\dots,I\}$. The $i$-th UE can be associated with the MEC and MCC servers to offload the computation of the machine translation tasks. Particularly, the end devices can execute the tasks in three modes simultaneously, including local execution, offloading the tasks to the MEC server, and offloading the tasks to the MCC server, which is further illustrated in~\ref{chp6:queue}. Denote the association parameter to the MEC server and MCC server as $\rho^{\rm{E}}_i \in \{0, 1\}$ and $\rho^{\rm{C}}_i \in \{0, 1\}$, respectively. For the local execution mode, i.e., $\rho^{\rm{E}}_i = \rho^{\rm{C}}_i = 0$ the end devices execute the tasks by the local models, which leads to the highest energy consumption and executing latency, but can guarantee the task execution performance when the channel state is bad. The MCC offloading mode is with $\rho^{\rm{C}}_i = 1$, where the raw source data is offloaded to the MCC servers directly, which saves the execution costs at the end devices but leads to increased transmission overhead and high transmission latency due to the restricted access bandwidth to the centralized BS. For the MEC offloading mode with $\rho^{\rm{E}}_i = 1$, on the other hand, only offloads the extracted task-related semantic information to the MEC servers and executes part of the tasks locally, hence reducing the transmission overhead and making a tradeoff between the transmission cost and computation cost. 

\subsection{Queuing Model}

Without loss of generality, we assume that the average number of tasks arriving at the $i$-th UE are $\Lambda_i$, which is independent and identically distributed (i.i.d) over time slots and follows the Poisson distribution.

The queue length of $U_i$ at the beginning of time slot $t$ can be expressed as $Q^{\rm {L}}_i(t)$, then the queue update process will be expressed as 
\begin{equation}
    Q^{\rm {L}}_i(t+1) = \mathrm{max}
(Q^{\rm {L}}_i(t) - \mu^{\rm {L}}_i(t), 0) + \Lambda_i(t),
\label{chp6:local_queue}
\end{equation}
where $\mu^{\rm {L}}_i(t) = u^{\rm {L}}_i(t) + u^{\rm{E}}_i(t) + u^{\rm{C}}_i(t)$ denotes the local execution rate, $u^{\rm {L}}_i(t)$ represents the number of tasks finished by the local execution, $u^{\rm{E}}_i(t)$ represents the number of tasks that offloaded to the MEC server, and $u^{\rm{C}}_i(t)$ represents the number of tasks offloaded to the MCC server that the source information is transmitted during the time slot $t$.

Meanwhile, we suppose that the MEC server maintains $I$ dedicated queues for each UE, and the queue length of the MEC server at the beginning of the time slot $t$ can be derived by 
\begin{equation}
    Q^{\rm{E}}_i(t+1) = \mathrm{max}
(Q^{\rm{E}}_i(t) - \mu^{\rm{E}}_i(t), 0) + u^{\rm{E}}_i(t),
\label{chp6:ES_queue}
\end{equation}
where $\mu^E_i(t)$ denotes the execution rate of the MEC server. The MCC server is assumed to have redundant computational resources and an adequate power supply, hence we do not consider the queue of the MCC server and the power consumption in optimization, which means the task offloaded to the MCC server can be executed with neglected latency and energy consumption.



In this paper, we focus on the computational resource system and assume that the task queues have a sufficiently large capacity and initially with empty inquiries, i.e., $Q^{\rm {L}}_i(0) = Q^{\rm{E}}_i(0) = 0, \forall i \in I$. According to~\cite{ROSS2014481}, the average waiting time experienced by UEs exhibits a direct proportionality to the average queue length, denoting the average quantity of tasks present within the system. Consequently, we utilize the long-term average queue length observed at both UEs and MEC as a metric for estimating task completion delays~\cite{10102429}. To guarantee the average latency and the QoS of the proposed cloud-edge-end network, constraints are set to restrict the average queue length of the UEs and ES as follows
\begin{equation}
    \lim_{T\to\infty}\frac{1}{T}\sum^T_{t=0}Q^{\rm{L}}_i(t) \leq  {Q^{\rm{L}}_{\max}},
\label{chp6:queue_max_L}
\end{equation}
and 
\begin{equation}
    \lim_{T\to\infty}\frac{1}{T}\sum^T_{t=0}Q^{\rm{E}}_i(t) \leq  {Q^{\rm{E}}_{\max}},
\label{chp6:queue_max_E}
\end{equation}
respectively. {It is noted that the long-term average backlog constraints are set to guarantee the task latency and QoS, but the hardware of users and MEC can support the sufficient capacity of the queues at each time step.} To achieve this together with (\ref{chp6:local_queue}), we also guarantee that $ {Q^{\rm{L}}_{\max}} \geq \Lambda_i$.

Generally, the task we investigate in this paper of the end devices can be executed by the local computation, the edge servers, and the cloud servers. To build up the task-executing model, we first define the task-executing related parameters. 

\subsection{Task Execution Model}
In this subsection, we investigate the task execution progress during a time slot with the time slot length $\tau$, and hence we omit the number of time slots $t$ as the model is the same in each time slot. Assuming that the models that we applied to execute the data are identical at the end devices, the MEC server, and the MCC server. The execution of the task includes the encoding process to help the machine understand the task, and the decoding process to generate the results for the task, with the hardware requirement $l = l^{\rm{en}} + l^{\rm{de}}$, respectively.

For the locally executed modes, the number of tasks that can be executed within $\tau$ of $U_i$ can be expressed by 
\begin{equation}
    u_i^{\rm{L}} = \tau n^{\rm{L}}_i f^{\rm{L}}_i / l,
\end{equation}
where $f^{\rm{L}}_i$ is the GPU clock frequency of the end devices, and $n^{\rm{L}}_i$ is the number of floating-point operations that the GPU can execute per cycle of the end device $U_i$. Similarly, for the task offloaded to the MEC server, the end devices need to encode the tasks, and the encoding volume can be expressed by 
\begin{equation}
    u^{\rm{E}}_i= \tau n^{\rm{L}}_i f^{\rm{en}}_i / l^{\rm{en}},
\end{equation}
{where $f^{\rm{en}}_i$ is the clock frequency for encoding the raw tasks}, and the execution power for the end device $U_i$ is given by $p^{\rm{L}}_i = \alpha^{\rm{L}} ({f^{\rm{L}}}_i)^3 + \alpha^{\rm{L}} ({f^{\rm{en}}}_i)^3$ according to circuit theory, where $\alpha^{\rm{L}}$ is the local power coefficient which depends on the hardware architecture. Then, the semantic features are offloaded to MEC servers, where the features are further decoded to the extracted task-related information and the tasks are completed. The source task execution rate and the semantic decoding volume can be similarly denoted as 
\begin{equation}
    \mu^E_i = \tau n^{\rm{E}}_i f^{\rm{E}}_i / l^{\rm{de}},
\end{equation}
where $f^{\rm{E}}_i$ is the GPU clock frequency, and $n^{\rm{E}}_i$ is the floating-point operations that the GPU of the MEC server can execute per cycle. {The power consumption of the MEC server for $U_i$ can be similarly defined by $p^{\rm{E}}_i = \eta^{\rm{E}} \alpha^{\rm{E}} ({f^{\rm{E}}}_i)^3$, where $\alpha^{\rm{E}}$ is the power coefficient of the MEC server, which maps the computation frequency to the weighted power consumption.}\footnote{The value of the power at the end devices and the MEC server is different, and {a weight coefficient $\eta^{\rm{E}}$} is added to $\alpha^{\rm{E}}$ to evaluate the power consumption of the system more fairly.} It is noted that the computational capacity of MEC is limited by the maximum GPU frequency, i.e., $f^{\rm{E}}_i \leq f^{\rm{E}}_{\max}$. Meanwhile, due to the transmission signaling overhead and bandwidth restriction, we assume that the MEC server and MCC server can serve at most $\chi^{\rm{E}}_{\max}$ and $\chi^{\rm{C}}_{\max}$ end devices, respectively.

\subsection{Computational Offloading Model}
Consider the cloud-edge-end network with MEC and MCC servers is structured on the orthogonal frequency-division multiple access (OFDMA) scheme. The MEC and MCC servers can communicate with the $U_i$ on the bandwidth $b^{\rm{E}}_i$ and $b^{\rm{C}}_i$. Suppose that the channels consist of large-scale fading and small-scale fading, and the SNR between $U_i$ and MEC and MCC servers can be denoted as 

\begin{equation}
    \gamma^{\rm{E}}_{i} = \frac{{p^{\rm{tx, E}}_i}|h^{\rm{E}}_i|^2}{b^{\rm{E}}_i \sigma^2},
\end{equation}
and
\begin{equation}
    \gamma^{\rm{C}}_{i} = \frac{{p^{\rm{tx, C}}_i}|h^{\rm{C}_i}|^2}{b^{\rm{C}}_i \sigma^2},
\end{equation}
respectively, where $ p^{\rm{tx,E}}_i$ and $p^{\rm{tx,C}}_i$ are the transmit power, $h^{\rm{E}}_i = \sqrt {g^{\rm{E}}_i} \widetilde h^{\rm{E}}_i$ and $h^{\rm{C}}_i = \sqrt {g^{\rm{C}}_i\psi} \widetilde h^{\rm{C}}_i$ represent channel coefficients to the MEC server and MCC server, {$\widetilde h^{\rm{E}}_i$ and $\widetilde h^{\rm{C}}_i$ are the Rician fading coefficient to the MEC server and Rayleigh fading coefficient to MCC server}, respectively, $\psi$ denotes the shadowing effect, and $\sigma^2$ is the noise power spectral density.

In the traditional Shannon paradigm, the data rate is measured by the bit rate. On the other hand, the transmission rate of semantic communication is based on the semantic unit, and the semantic rate should consider the amount of semantic information. The input source data of the semantic encoder is denoted as $\boldsymbol{x_i}=\{x[0], x[1], \dots, x[S_i]\}$, where $S_i$ represents the length of the input data of $U_i$. For example, for a text semantic encoder, the input $\boldsymbol{x_i}$ is a sentence, $x[\cdot]$ represents the words in the input sentence, while $S_i$ is the number of words in a sentence. The computational hardware requirement for each task can be calculated in the same way as Appendix~\ref{appendix_1}.

The semantic offloaded rate can be expressed by 
\begin{equation}
    \Gamma_i = \frac {b_i \epsilon_i}{S_ik_i},
\end{equation}
where $\epsilon_i \sim f(k_i, \gamma_i)$ is the task execution accuracy that is related to the $k_i$ and the SNR $\gamma_i$, $k_i$ denotes the average number of semantic symbols, i.e., the output dimension of the semantic encoders, used for each input unit $x[\cdot]$. Note that $S_i$, and $k_i$ are the parameters that depend on the pre-trained DeepSC model and the source type, and can be considered constant values during the resource optimization process. {Given that the end users need to offload the encoded semantic features of volume $u^{\rm{E}}_i = \Gamma_i \tau $ to the MES server over a given bandwidth $b_i$}, the accuracy should satisfy 
\begin{equation}
    \epsilon_i = \frac{u^{\rm{E}}_iS_ik_i}{\tau b_i},
\end{equation}
which further restricted the transmit power $p^{\rm{tx, E}}_i$. For the tasks offloaded to the MCC server, the source data is mapped into bits through the source coding. According to Shannon's Theory, the transmit power for offloading to the MCC server should satisfy that

\begin{equation}
    p^{\rm{tx, C}}_i = \left(2^{\frac{v^{\rm{C}}_iS_ik_0}{\tau b_i}}\right)\frac{\sigma^2b_i}{|{h^{\rm{C}}_i}|^2},
\end{equation}
where $k_0$ represents the average bit used for encoding a word, and $v^{\rm{C}}_iS_ik_0$ represents the total bits that need to be transmitted for the offloading volume $v^{\rm{C}}_i$. 

We assume that the bandwidth is equally allocated to the associated UEs and is sufficient for supporting the offloading process, i.e., $b^{\rm{E}}_i = \frac{BW^{\rm{E}}}{\chi^{\rm{E}}}$ and $b^{\rm{C}}_i = \frac{BW^{\rm{C}}}{\chi^{\rm{C}}}$, where the sum bandwidth of the MEC and MCC servers are $BW^{\rm{E}}$ and $BW^{\rm{C}}$, respectively.

\section{The Multi-Stage MINLP Problem of Energy Consumption Minimization}
\label{chp6:transform}

\subsection{Problem Formulation} 
In this paper, we aim to minimize the energy consumption of the UEs and the MEC server for task execution, with constraints on the stability of task queues and the limitations on the computational and communication resources. Accordingly, the energy consumption in time slot $t$ can be calculated as\footnote{Note that the static and circuit power for maintaining the basic operations can be considered as constant values, which can be put aside from the objective function.}

\begin{equation}
    p(t) = \sum^I_{i = 0} p^{\rm{L}}_i(t) + p^{\rm{E}}_i(t) + p^{\rm{tx, C}}_i(t) + p^{\rm{tx, E}}_i(t).
\end{equation}

Due to the unstable CSI and the unpredictable upcoming tasks, we focus on the long-term average energy consumption of the system. The optimization variables include the bandwidth allocation, local computational frequency, local offloading policy, and offloading power, which can be concatenated as $\boldsymbol{X} = \{\boldsymbol{X}(t)\}_{t \in T}$. The optimization parameters in time slot can be denoted as $\boldsymbol{X}(t) = \{ \boldsymbol{\rho}(t), \boldsymbol{f}(t),\boldsymbol{u}(t)\}_{t\in T}$, denoting the combined vector of variables at time slot $t$: $\boldsymbol{\rho}(t) = \{(\rho^{\rm{E}}_i(t), \rho^{\rm{C}}_i(t))_{i \in I}\}$, $\boldsymbol{f}(t) = \{(f^{\rm{L}}_i(t),f^{\rm{E}}_i(t))_{i \in I}\}$, $\boldsymbol{u}(t) = \{(u^{\rm{E}}_i(t), u^{\rm{C}}_i(t))_{i \in I}\}$. Then, a multi-stage MINLP problem can be formulated as

\begin{mini!}|l|
{\boldsymbol{X}}{\lim_{t\rightarrow \infty}\frac{1}{T}\sum_{t=0}\mathbb{E}\left[p(t)\right]}
{\label{chp6:eq20}}{(\textbf{P1})\medspace}
\addConstraint{\rho^{\rm{E}}_i(t) \in \{0, 1\}, i \in I\label{chp6:objective:c8} }
\addConstraint{\rho^{\rm{C}}_i(t) \in \{0, 1\}, i \in I\label{chp6:objective:c9} }
\addConstraint{\sum^I_{i = 0}\rho^{\rm{E}}_i(t) \leq \chi^{\rm{E}}_{\max}\label{chp6:objective:c10} }
\addConstraint{\sum^I_{i = 0}\rho^{\rm{C}}_i(t) \leq \chi^{\rm{C}}_{\max}\label{chp6:objective:c11} }
\addConstraint{\epsilon_i(t) \geq \epsilon_{min}\label{chp6:objective:c1} }
\addConstraint{f^{\rm{L}}_i(t) + f^{\rm{en}}_i(t) \leq f^{\rm{L}}_{\max} \label{chp6:objective:c2} 
}
\addConstraint{f^{\rm{E}}_i(t)\leq f^{\rm{E}}_{\max}\label{chp6:objective:c3}
}
\addConstraint{u^{\rm {L}}_i(t) + u^{\rm{E}}_i(t) + u^{\rm{C}}_i(t)\leq Q^{\rm {L}}_i(t)\label{chp6:objective:c4} 
}
\addConstraint{\mu^E_i(t) \leq Q^{\rm{E}}_i(t)\label{chp6:objective:c5} 
}
\addConstraint{\lim_{t\rightarrow \infty} \frac{\mathbb{E}[Q^{\rm {L}}_i(t)]}{t}\leq Q^{\rm {L}}_{\max}\label{chp6:objective:c6} 
}
\addConstraint{\lim_{t\rightarrow \infty} \frac{\mathbb{E}[Q^{\rm {E}}_i(t)]}{t} \leq Q^{\rm {E}}_{\max}.\label{chp6:objective:c7} 
}
\end{mini!}
In (\textbf{P1}), (\ref{chp6:objective:c8}) and (\ref{chp6:objective:c9}) denote the mode selection constraints of the end devices, and (\ref{chp6:objective:c10}) and (\ref{chp6:objective:c11}) further denote the maximum access restriction on the MEC and MCC server. Additionally, (\ref{chp6:objective:c1}) guarantees the task execution accuracy for the semantic-aware offloading, (\ref{chp6:objective:c2}) restrict the maximum local computational frequency, 
(\ref{chp6:objective:c3}) guarantee the computational frequency does not exceed the total capacity of the MEC server. Meanwhile, (\ref{chp6:objective:c4}) and (\ref{chp6:objective:c5}) guarantee that only the maximum amount of task queues backlog can be processed within a time slot. Finally, (\ref{chp6:objective:c6}) and (\ref{chp6:objective:c7}) are set according to (\ref{chp6:queue_max_L}) and (\ref{chp6:queue_max_E}) to guarantee the QoS performance of the task execution.

It is observed that (\textbf{P1}) is a stochastic optimization problem of a time-evolving system. The resource management and the offloading policy need to be made under the stochastic channels and data arrivals. Moreover, solving (\textbf{P1}) is challenging because of the coupled variables and temporally correlated resource management, which means the over-aggressive approach may not solve the problem effectively~\cite{10102429}. The fast-varying channels also demand a low-complexity algorithm to realize real-time policy optimization. In such conditions, we propose a hybrid algorithm to solve the formulated (\textbf{P1}). Specifically, we first adopt the Lyapunov optimization framework to decouple the long-term optimization problem into per-frame deterministic problems, and apply the model-free reinforcement learning framework to provide a low-complexity sub-optimal solution. 

\subsection{Lyapunov-Guided Problem Transformation}
In this section, the Lyapunov optimization is adopted to decouple (\textbf{P1}) into deterministic problems in per time slot. First, to cope with the queue length constraints (\ref{chp6:objective:c6}) and (\ref{chp6:objective:c7}), we introduce two virtual queues for each end device and MEC as~\cite{1941130}

\begin{equation}
    Z^{\rm{L}}_i(t+1) = \max\left\{Z^{\rm{L}}_i(t) + Q^{\rm{L}}_i(t+1) - Q^{\rm {L}}_{\max}, 0\right\},
\label{chp6:v_queue}
\end{equation}
and
\begin{equation}
    Z^{\rm{E}}_i(t+1) = \max\left\{Z^{\rm{E}}_i(t) + Q^{\rm{E}}_i(t+1) - Q^{\rm {E}}_{\max}, 0\right\},
\end{equation}
respectively, where $Z^{\rm{L}}_i(0) = Z^{\rm{E}}_i(t) = 0$. The constraints (\ref{chp6:objective:c6}) and (\ref{chp6:objective:c7}) can be satisfied if $\lim_{t\rightarrow \infty} \frac{\mathbb{E}[Z^{\rm {L}}_i(t)]}{t} = 0$ and $\lim_{t\rightarrow \infty} \frac{\mathbb{E}[Z^{\rm {E}}_i(t)]}{t} = 0$. The virtual queue $Z^{\rm {L}}_i(t)$ can be understood as a queue with random "arrivals" $Q^{\rm{L}}_i(t+1)$ and maximum "processing rate" $Q^{\rm {L}}_{\max}$. It is obvious that the real queue length $Q^{\rm{L}}_i(t+1)$ does not exceed $Q^{\rm {L}}_{\max}$ if the length of the virtual queues is stable, and thus (\ref{chp6:objective:c6}) is satisfied.

To jointly consider the real queue length and the virtual queue to guarantee the QoS requirement, we define the total queue backlog as $\boldsymbol{\Theta}(t) =  \left\{(Q^{\rm{L}}_i(t), Q^{\rm{E}}_i(t), Z^{\rm{L}}_i(t), Z^{\rm{E}}_i(t))_{i \in I}\right\}$. To measure the total queue backlog and consider the stability of the queues, we define the Lyapunov function and the Lyapunov drift function as
\begin{equation}
    {\mathcal{L}}(\boldsymbol{\Theta}(t)) = \frac{1}{2}\left[(Q^{\rm{L}}_i(t))^2 + (Q^{\rm{E}}_i(t))^2 + (Z^{\rm{L}}_i(t))^2 + (Z^{\rm{E}}_i(t))^2\right],
\label{chp6:Lyapunov_function}
\end{equation}
and
\begin{equation}
    \Delta{\mathcal{L}}(\boldsymbol{\Theta}(t)) = {\mathbb E}\left[{\mathcal{L}}(\boldsymbol{\Theta}(t+1)) - {\mathcal{L}}(\boldsymbol{\Theta}(t))|\boldsymbol{\Theta}(t)\right].
\label{chp6:Lyapunov_drift}
\end{equation}

{To minimize the long-term average power consumption of the system, while ensuring the queue stability constraint, we define the Lyapunov-drift-plus-penalty as}
\begin{equation}
    \Delta_v{\mathcal{L}}(\boldsymbol{\Theta}(t)) = \Delta{\mathcal{L}}(\boldsymbol{\Theta}(t)) + v\mathbb{E}\left[p(t)|\boldsymbol{\Theta}(t)\right],
\label{chp6:Lyapunov_drift_penalty}
\end{equation}
where $v$ is a positive value that represents the preference weight on the average queuing delay and the power consumption.

To transform the long-term multi-stage problem (\textbf{P1}) into a deterministic problem, we employ the opportunistic expectation minimization technique as outlined in~\cite{1941130}. Specifically, the objective of long-term power minimization involves minimizing the upper bound of~(\ref{chp6:Lyapunov_drift}) within each time frame with the observed total queue backlog $\boldsymbol{\Theta}(t)$, which is provided and proved in \textbf{Theorem 1}.

\begin{theorem}[Upper bound of the Lyapunov drift function]
The Lyapunov-drift-plus-penalty $\Delta_v{\mathcal{L}}(\boldsymbol{\Theta}(t))$ is bounded as
\begin{equation}
\begin{split}
    \Delta_v{\mathcal{L}}(\boldsymbol{\Theta}(t)) \leq &\hat{B}  - \sum_{i \in I} \mathbb{E}[(Q^{\rm{L}}_i(t) + Z^{\rm{L}}_i(t))(\mu^{\rm{L}}_i(t) - \Lambda_i(t)) \\
    &- (Q^{\rm{E}}_i(t) + Z^{\rm{E}}_i(t))(\mu^{\rm{E}}_i(t) - u^{\rm{E}}_i(t))|\boldsymbol{\Theta}(t)] \\
    & + v\mathbb{E}\left[p(t)|\boldsymbol{\Theta}(t)\right],
\label{chp6:Lyapunov_upper}
\end{split}
\end{equation}
where $\hat{B}$ can be removed for the optimization of the target variable $\boldsymbol{X}(t)$ as it consists of constant and independent terms from the observation.
\begin{proof}
Please refer to Appendix \ref{appendix_1}.
\end{proof}
\label{chp6:theorem_1}
\end{theorem}

Based on the aforementioned analysis, the deterministic problem within each time slot is formulated as
\begin{mini!}|l|
{\boldsymbol{X}(t)}{G(\boldsymbol{X}(t))}
{\label{chp6:P1}}{\textbf{(P2)}\medspace}
\addConstraint (\ref{chp6:objective:c8}) - (\ref{chp6:objective:c7}),
\end{mini!}
with the objective function
\begin{equation}
\begin{split}
    &G(\boldsymbol{X}(t))  = - \sum^I_{i = 0} \left[(Q^{\rm{L}}_i(t) + Z^{\rm{L}}_i(t))(\mu^{\rm{L}}_i(t) - \Lambda_i(t))\right] \\
    & - \sum^I_{i = 0} \left[(Q^{\rm{E}}_i(t) + Z^{\rm{E}}_i(t))(\mu^{\rm{E}}_i(t) - u^{\rm{E}}_i(t))\right] \\
    & + v\left(\sum^I_{i = 0} p^{\rm{L}}_i(t) + p^{\rm{E}}_i(t) + p^{\rm{tx, C}}_i(t) + p^{\rm{tx, E}}_i(t)\right).
\label{chp6:Lyapunov_objective}
\end{split}
\end{equation}
We can observe that (\textbf{P2}) only requires the current queue and channel state information, and does not rely on any future information about the incoming tasks and channel, which makes (\textbf{P2}) a Markov process that can be solved by the online optimization approach. However, the difficulty still lies in solving the MINLP in each time slot. 

Denoting that the variables $\boldsymbol{X}(t) = \{\boldsymbol{\rho}(t), \boldsymbol{y}(t)\}$, where $\boldsymbol{\rho}(t)$ is the binary offloading policy and $\boldsymbol{y}(t) = \{\boldsymbol{f}(t),\boldsymbol{u}(t)\}$ is the continuous variables to control the computational resource allocation. It is observed that given the offloading policy $\boldsymbol{\rho}(t)$, (\textbf{P2}) becomes the convex problem and can be solved with the low-complexity convex optimization algorithms. 

To find the optimal offloading policy, one way is to use the exhaustive search algorithm to find all possible offloading combinations, which demands significant computational complexity and is hard to complete within the short time slot. A low-complexity algorithm to address the Markov problem is the reinforcement learning algorithm, and we also combine the model-free reinforcement learning algorithm with the model-based online convex optimization algorithm, which is introduced in Section~\ref{chp6:DRLH}.

\section{Deep Reinforcement Learning based Hybrid Framework}
\label{chp6:DRLH}
\begin{figure*}[t]
\centering
\includegraphics[width=1.8\columnwidth]{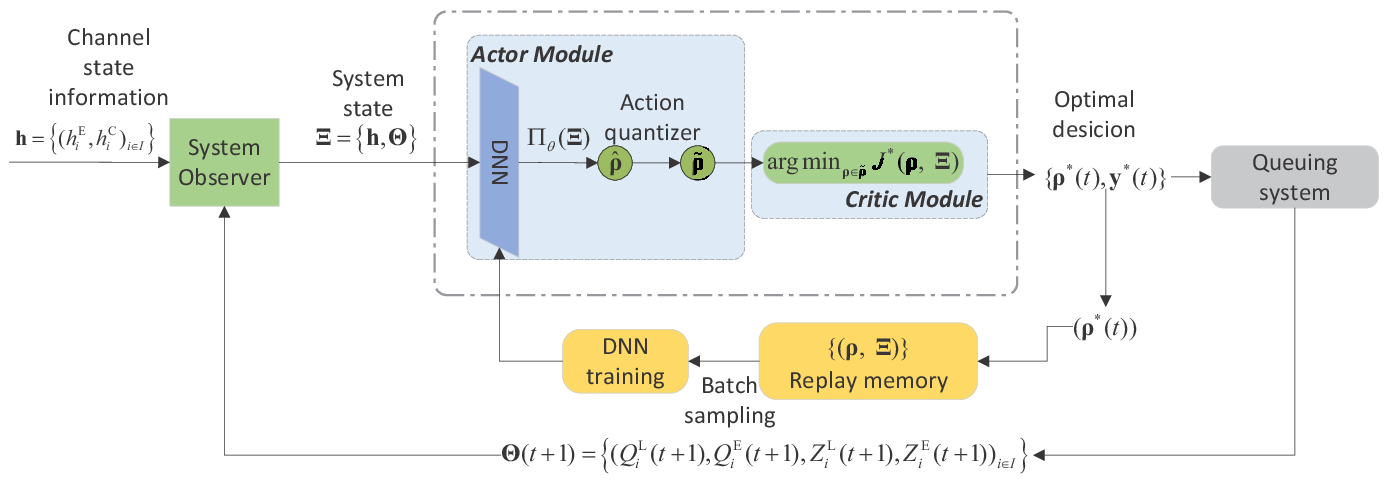}
\caption{The proposed \textit{DRLH} framework.}
\label{chp6:framework}
\end{figure*}

The proposed (\textit{DRLH}) framework is illustrated as Fig.~\ref{chp6:framework}. Particularly, the \textit{DRLH} framework consists of three components, including an actor module, a critic module, and a policy update module. As (\textbf{P2}) is the deterministic problem in each time slot, for simplicity of exposition, we drop the number of the time slot $t$. The actor model forms several potential decisions for the policy choice $\boldsymbol{\rho}= \left\{(\rho^{\rm{E}}_i, \rho^{\rm{C}}_i)_{i \in I}\right\}$. {For the exhaustive search algorithm, all of the possible actions are evaluated by the critic module to generate the optimal resource allocation policy. On the other hand, only the candidate policies are evaluated by solving the remaining variables with the model-based optimization approach in the proposed \textit{DRLH} framework, which reducing the operation times of the model-based optimization of the critic module significantly.} Finally, the optimal decision is made and performed in the online queue system, and the optimal solution for the policy action and the system state is stored in the replay memory for training the DNN in the actor module. In the following section, we introduce the proposed framework and the modules in detail.

\subsection{\textit{DRLH} Components}
\subsubsection{Model-free actor module}
Collecting the channel and the total queue backlog information as the input of the DNN, and processing the action quantizer module, the actor module outputs several candidate choices for the offloading policy. At the beginning of each time slot, the DNN takes the system state $\boldsymbol{\Xi} = \left\{\boldsymbol{h},\boldsymbol{\Theta} \right\}$ as the input, where $\boldsymbol{h} = \left\{(h^{\rm{E}}_i, h^{\rm{C}}_i)_{i \in I}\right\}$ represents the channel coefficient information at time slot $t$. The DNN then generates $N_{\rm{A}}$ candidate offloading policies $\boldsymbol{\hat{\rho}} = \{\boldsymbol{\hat{\rho}}^{n_A}\}^{N_{\rm{A}}}_{n_A=1}$. Particularly, each policy $\boldsymbol{\hat{\rho}}^{n_A} = \{\hat{\rho}^{n_A}_i \in [0,1]\}_{i \in I}$. Meanwhile, as the offloading policy is discrete, we adopt an action quantizer module after the DNN to convert the continuous output of DNN to discrete offloading policy $\boldsymbol{\widetilde{\rho}} = \{\boldsymbol{\widetilde{\rho}}^{n_A}\}^{N_{\rm{A}}}_{n_A=1}$ with $\boldsymbol{\widetilde{\rho}}^{n_A} = \{\widetilde{\rho}^{n_A}_i \in \{0,1\}\}_{i \in I}$.

Specifically, we use the DNN to approximate the relaxed offloading policy as $\Pi_\theta: \boldsymbol{\Xi} \rightarrow \boldsymbol{\hat{\rho}}$. To ensure that the output policy $\hat{\rho}^{n_A}_i \in [0, 1], i \in I$, we apply the sigmoid activation function at the output layer of the DNN. Then, the quantizer algorithm convert the relaxed offloading policy $\boldsymbol{\hat{\rho}} \rightarrow \boldsymbol{\widetilde{\rho}}$. The idea is to find the value $\upsilon_{\rm{th}}$ for the $\chi_{\max}$-th largest element in the relaxed policy $\boldsymbol{\hat{\rho}}$. Then we set the elements which is larger than or equal to the value for $\upsilon_{\rm{th}}$ to 1, and others to 0.

The DNN is trained dynamically and updated periodically adapt to cope with the time-varying channel states. Finally, it should be able to approximate an optimal policy $\Pi^*$, which maps the input system state $\boldsymbol{\Xi} \rightarrow (\boldsymbol{\widetilde{\rho}})^*$. 

\subsubsection{Model-based critic model}
The critic module evaluates the $N_{\rm{A}}$ candidate offloading policies from the actor module and selects the best offloading action $\boldsymbol{\rho}$ as
\begin{equation}
    \boldsymbol{\rho} = \arg\min_{\boldsymbol{\rho} \in \boldsymbol{\widetilde{\rho}}} J^*( \boldsymbol{\rho}, \boldsymbol{\Xi}).
\end{equation}
It is noted that the calculation of the $J^*( \boldsymbol{\rho}, \boldsymbol{\Xi})$ is performed by $N_{\rm{A}}$ times, hence we need to consider the performance-complexity tradeoff by setting the hyper-parameter $N_{\rm{A}}$ properly.

\subsubsection{Policy update module}
The function of the policy update module to train the model-free DNN is to store the system state and optimal policy pair 
$\{(\boldsymbol{\rho}, \boldsymbol{\Xi})\}$ to the replay memory and use the cross-entropy as the loss function. By minimizing the difference between the policy of the DNN $\Pi_\theta$ and the optimal offloading policy $\rho$ evaluated by the critic module, the DNN of the actor module can generate the actions that are closer to the optimal action. The cross-entropy loss function can be expressed by

\begin{equation}
\begin{split}
    L(\theta) = &-\frac{1}{|{\cal B}|}\sum_{\tau \in {\cal B}}[(\boldsymbol{\rho}(\tau)    )^\top\log(\Pi_\theta\boldsymbol{\Xi}(\tau)) \\ &+ (1 - \boldsymbol{\rho}(\tau)) ^\top \log (1 - \Pi_\theta\boldsymbol{\Xi}(\tau))],
\end{split}
\end{equation}
where $\tau$ represents the sampled time steps from the replay memory $\cal {B}$.
\subsection{Model-Based Optimization of the Critic Module}
In this section, we propose a low-complexity algorithm to obtain $J^*(\boldsymbol{\rho}, \boldsymbol{\Xi})$ with the {given $(\boldsymbol{\rho}, \boldsymbol{\Xi})$ pairs}.

\begin{equation}
\begin{aligned} \label{chp6:P2}
(\textbf{P3}) \medspace \min_{\boldsymbol{f},\boldsymbol{u}} \quad &- \sum^I_{i = 0} G(\boldsymbol{f},\boldsymbol{u})\\
&\begin{array}{r@{\quad}r@{}l@{\quad}l}
s.t. & (\ref{chp6:objective:c1}) - (\ref{chp6:objective:c7}). \\
\end{array}
\end{aligned}
\end{equation}

We further decompose the (\textbf{P3}) into sub-problems, including the optimization for the volume of offloading for the MEC server $\boldsymbol{u}^{\rm{E}}$ and MCC server $\boldsymbol{u}^{\rm{C}}$, and optimization for the computation capacity for the end devices $\boldsymbol{f}^{\rm{L}}$ and for the MEC server $\boldsymbol{f}^{\rm{E}}$. 

\subsubsection{Optimization on offloading volume for MEC server}
The sub-problem for optimization for the offloading volume for MEC Server can be formulated as
\begin{equation}
\begin{aligned} \label{chp6:P6_3_1}
(\textbf{P3.1}) \medspace \min_{\boldsymbol{u}^{\rm{E}}} \quad &- \sum^I_{i = 0} \left[(Q^{\rm{L}}_i+ Z^{\rm{L}}_i - Q^{\rm{E}}_i - Z^{\rm{E}}_i)(u^{\rm{E}}_i)\right] \\
 &+ v\left(\sum^I_{i = 0}  \alpha^{\rm{L}}(f^{\rm{en}}_i)^3 + p^{\rm{tx, E}}_i\right)\\
&\begin{array}{r@{\quad}r@{}l@{\quad}l}
s.t. & 0 \leq u^{\rm{E}}_i \leq \min\left\{Q^{\rm{L}}_i, u^{\rm{E}}_{\max}\right\},\\
\end{array}
\end{aligned}
\end{equation}
where $u^{\rm{E}}_{\max} = \tau \frac{f^{\rm{L}}_{\max}n^{\rm{L}}_i}{l^{\rm{en}}}$ denotes the upper bound of the $u^{\rm{E}}_i$ with the maximum computational capability for the local encoding process. It is also noted that for the end devices that are not associated with the MEC server, i.e., $\rho^{\rm{E}}_i = 0$, the offloading volume is 0, i.e., $u^{\rm{E}}_i = 0$.

The optimal resolution for the aforementioned problem corresponds to either the stationary point of (\ref{chp6:P6_3_1}) or one of its boundary points. Specifically, $(u^{\rm{E}}_i)^* = 0$ if $Q^{\rm{L}}_i+ Z^{\rm{L}}_i - Q^{\rm{E}}_i - Z^{\rm{E}}_i \leq 0$. Otherwise, $(u^{\rm{E}}_i)^* = \max\left\{\min\left\{\overline{u^{\rm{E}}_i},u^{\rm{E}}_{\max} \right\}\right\}$, where $\overline{u^{\rm{E}}_i}$ is the stationary point.

To acquire the value of $\overline{u^{\rm{E}}_i}$, we need to calculate the derivative of (\ref{chp6:P6_3_1}). However, it is hard to find the closed-form derivative for the transmit power $p^{\rm{tx, E}}_i$ as it is provided by the pre-trained DeepSC model~\cite{9252948} as a black box. Considering that the transmit power is much less than the computational power of the end devices, we omit the derivative item of the $p^{\rm{tx, E}}_i$, and apply the derivative items of queue length and the computational capacity to calculate the stationary point approximately. Hence, the stationary point for (\ref{chp6:P6_3_1}) can be denoted by $\overline{u^{\rm{E}}_i} = \sqrt{\left(\frac{\tau n^{\rm{L}}_i}{l^{\rm{en}}_i}\right)^3\frac{Q^{\rm{L}}_i+ Z^{\rm{L}}_i - Q^{\rm{E}}_i - Z^{\rm{E}}_i}{3v\alpha^{\rm{L}}}}$.

\subsubsection{Optimization on offloading volume for MCC server}
Similarly, the sub-problem for the optimization of the offloading volume for MCC server can be denoted as

\begin{equation}
\begin{aligned} \label{chp6:P6_3_2}
(\textbf{P3.2}) \medspace \min_{\boldsymbol{u}^{\rm{C}}} \quad &- \sum^I_{i = 0} \left[(Q^{\rm{L}}_i+ Z^{\rm{L}}_i - (u^{\rm{E}}_i)^*)(u^{\rm{C}}_i)\right] \\
 &+ v\left(\sum^I_{i = 0} p^{\rm{tx, C}}_i\right).\\
&\begin{array}{r@{\quad}r@{}l@{\quad}l}
s.t. & 0 \leq u^{\rm{C}}_i \leq \min\left\{Q^{\rm{L}}_i - (u^{\rm{E}}_i)^*, u^{\rm{C}}_{\max}\right\},\\
\end{array}
\end{aligned}
\end{equation}
where $u^{\rm{C}}_{\max} = \frac{\tau b^{\rm {C}}_i }{S_ik_0}\log_2 \left(1 + \frac{{p^{\rm{tx, C}}_{\max}}|{h^{\rm{C}}_i}|^2}{b^{\rm{C}}_i \sigma^2}\right)$, represents the offloading upper bound to the MCC server using the maximum transmit power. Similarly, the optimal solution to the problem (\textbf{P3.2}) corresponds to either the stationary point of (\ref{chp6:P6_3_2}) or one of the boundary points. Specifically, $(u^{\rm{C}}_i)^* = \max\left\{\min\left\{\overline{u^{\rm{C}}_i}, \min\left\{Q^{\rm{L}}_i - (u^{\rm{E}}_i)^*, u^{\rm{C}}_{\max}\right\}\right\}, 0\right\}$, where $\overline{u^{\rm{C}}_i} = \frac{\tau b^{\rm{C}}_i}{S_ik_0}\ \log_2 \left(\frac{(Q^{\rm{L}}_i+ Z^{\rm{L}}_i - (u^{\rm{E}}_i)^*)\tau|{h^{\rm{C}}_i}|^2}{\ln{2}\cdot vS_ik_0\sigma^2}\right)$ is the boundary point.

\subsubsection{Optimization on computational frequency of end devices}
After acquiring the optimal solution for the offloading volume for each end device, the optimization of the computational capacity for end devices can be decomposed for each computational frequency $f^{\rm{L}}_i$ as 

\begin{equation}
\begin{aligned} \label{chp6:P6_3_3}
(\textbf{P3.3}) \medspace \min_{\boldsymbol{f}^{\rm{L}}} \quad & \sum^I_{i = 0} \left[- \tau Q^{\rm{L}}_i f^{\rm{L}}_i n^{\rm{L}}_i/l + v \alpha^{\rm{L}}(f^{\rm{L}}_i)^3\right]\\
&\begin{array}{r@{\quad}r@{\quad}l@{\quad}l}
s.t. \quad 0 \leq f^{\rm{L}}_i \leq f^{\rm{L}}_{\max} - f^{\rm{en}}_i,\\
u^{\rm{L}}_i \leq Q^{\rm{L}}_i - (u^{\rm{E}}_i)^* - 
 (u^{\rm{C}}_i)^*.\\
\end{array}
\end{aligned}
\end{equation}

It can be observed that the (\textbf{P3.3}) is a convex problem since the objective function (\ref{chp6:P6_3_3}) is convex and all constraints are linear. (\textbf{P3.3}) can be further decomposed for each $f^{\rm{L}}_i$ and done by solving each $f^{\rm{L}}_i$. The optimal solution for (\textbf{P3.3}) corresponds to either the stationary point of the objective function (\ref{chp6:P6_3_3}) or one of the boundary points as $(f^{\rm{L}}_i)^* = \min\left\{\overline{f^{\rm{L}}_i},f^{\rm{L}}_{\max} - f^{\rm{en}}_i, \frac{(Q^{\rm{L}}_i - (u^{\rm{E}}_i)^* -  (u^{\rm{C}}_i)^*)l}{\tau n^{\rm{L}}_i} \right\}$, where $\overline{f^{\rm{L}}_i} = \sqrt{\frac{\tau Q^{\rm{L}}_i n^{\rm{L}}_i}{3vl\alpha^{\rm{L}}}}$ is the stationary point.

\subsubsection{Optimization on the computational frequency of MEC server}
The computational frequency of the MEC server can be optimized by solving (\textbf{P3.4}) as follows

\begin{equation}
\begin{aligned} \label{chp6:P6_3_4}
(\textbf{P3.4}) \medspace \min_{\boldsymbol{f}^{\rm{E}}} \quad & \sum^I_{i = 0} \left[- \tau Q^{\rm{E}}_i f^{\rm{E}}_i n^{\rm{E}}/l^{\rm{de}} + v \alpha^{\rm{E}}(f^{\rm{E}}_i)^3\right]\\
&\begin{array}{r@{\quad}r@{\quad}l@{\quad}l}
s.t. \quad 0 \leq f^{\rm{E}}_i \leq f^{\rm{E}}_{\max},\\
\mu^{\rm{E}}_i = \frac{\tau n^{\rm{E}} f^{\rm{E}}_i}{l^{\rm{de}}}\leq Q^{\rm{E}}_i.\\
\end{array}
\end{aligned}
\end{equation}
Similarly, (\textbf{P3.4}) is convex and can be decomposed into sub-problems for each $f^{\rm{E}}_i$. The optimal solution for each $f^{\rm{E}}_i$ is either the stationary point of (\ref{chp6:P6_3_4}) or one of the boundary points, i.e., $(f^{\rm{E}}_i)^* = \min\left\{\overline{f^{\rm{E}}_i},f^{\rm{E}}_{\max}, \frac{Q^{\rm{E}}_i l^{\rm{de}}}{\tau n^{\rm{E}}_i} \right\}$, where $\overline{f^{\rm{E}}_i} = \sqrt{\frac{\tau Q^{\rm{E}}_i n^{\rm{E}}_i}{3vl^{\rm{de}}\alpha^{\rm{E}}}}$ is the boundary point.

\begin{table}[tp]
\footnotesize
\begin{center}
\caption{Communication and Training Parameters}
\begin{tabular}{|c|c|}
\hline
\textbf{Parameter}&
\textbf{Value} \\ 
\hline
Time slot interval $\tau$&
$10$ ms \\
\hline
Maximum queue length of end devices $Q^{\rm{L}}_{\rm{max}}$&
$20$ tasks\\
\hline
Maximum queue length of MEC server $Q^{\rm{E}}_{\rm{max}}$&
$5$ tasks\\
\hline
The average length of the input data $S_i$&
$10$ words/sentence\\
\hline
The average number of symbols for each word $k_i$&
$24$ symbols/word \\
\hline
Computation coefficient of end devices $\alpha^{\rm{L}}$&
$5.787 \times 10^{-26}$\\
\hline
Weighted computation coefficient of MEC server $\alpha^{\rm{E}}$&
$4.45 \times 10^{-26}$\\
\hline
Maximum GPU frequency of end devices $f^{\rm{L}}_{\rm{max}}$&
$1.2\rm{GHz}$\\
\hline
Maximum GPU frequency of MEC server $f^{\rm{E}}_{\rm{max}}$&
$1.41\rm{GHz}$\\
\hline
Number of processors of the GPU at local devices&
2048\\
\hline
Number of processors of the GPU at MEC server&
6912\\
\hline
Minimum accuracy requirements $\epsilon_{min}$&
0.9\\
\hline
Training slots of the \textit{DRLH} framework&
15000\\
\hline
The value of the Lyapunov factor&
$2$\\
\hline
Learning rate of the \textit{DRLH} algorithm&
$1e^{-3}$\\
\hline
\end{tabular}
\label{chp6:tab1}
\end{center}
\end{table}

\section{Numerical Results}
\label{chp6:results}

In this section, the performance of the proposed DRLH framework is demonstrated by simulation results. We consider a hot spot with an MEC server deployed at the center, and $I=8$ end devices are randomly distributed in the hot spot within a radius from 50 to 150 meters. The BS with the MCC server is located 500 meters away from the hot spot. The maximum access devices to the MEC and MCC servers are set to $\chi^{\rm{E}}_{\rm{max}} = 4$ and $\chi^{\rm{C}}_{\rm{max}} = 2$, respectively. The other computational parameters and transmitted parameters are listed in Tabel~\ref{chp6:tab1} and Tabel~\ref{chp6:tab2}, respectively. 

\begin{table}[htp]
\footnotesize
\begin{center}
\caption{Transmission Parameters}
\begin{tabular}{|c|c|}
\hline
\textbf{Parameter}&
\textbf{Value} \\ 
\hline
Sum uplink bandwidth of MEC $BW^{\rm{E}}$&
1 $\rm{MHz}$\\
\hline
Sum uplink bandwidth of MCC $BW^{\rm{C}}$&
50 $\rm{kHz}$\\
\hline
noise power spectrum density&
-174 $\rm{dBm/Hz}$\\
\hline
Transmit power range&
$(0,20) \rm{dBm}$\\
\hline
Rician fading factor to the MEC server&
3dB \\
\hline
Rayleigh fading to the MCC server &
$\widetilde{h}^{\rm{C}}_i \thicksim \mathcal{CN}(0,1)$\\
\hline
\end{tabular}
\label{chp6:tab2}
\end{center}
\end{table}

\begin{figure}[t]
\centering
{
\includegraphics[width=0.8\columnwidth]{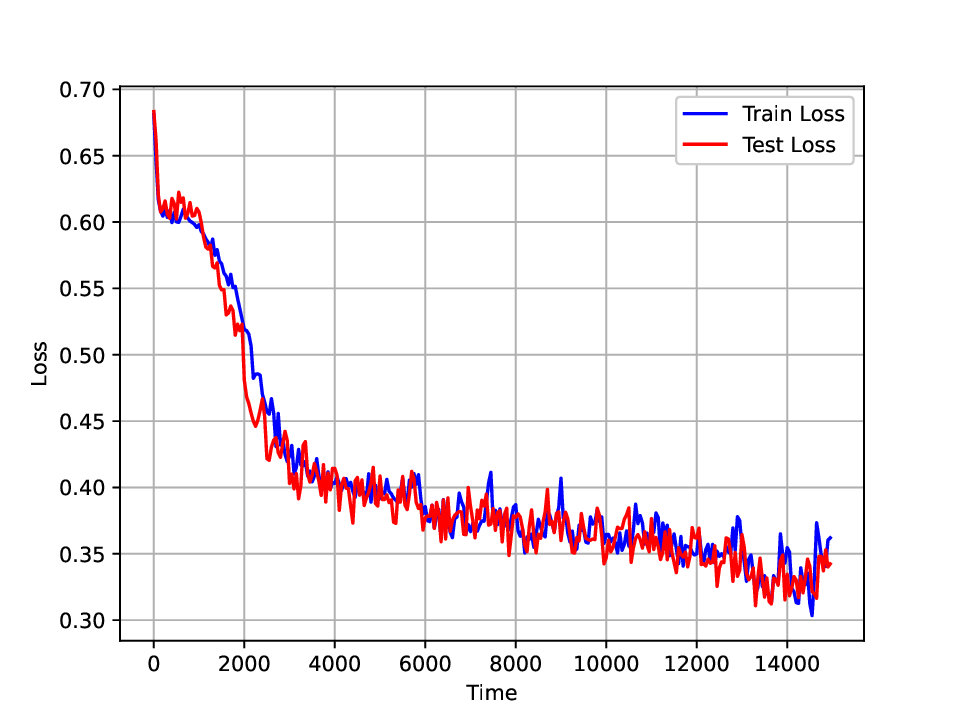}
}%
\caption{The training and testing loss the the proposed \textit{DRLH} framework.}
\label{chp6:loss}
\end{figure}

\begin{figure*}[t]
\centering
\subfigure[Queue length of users.]{
\includegraphics[width=0.66\columnwidth]{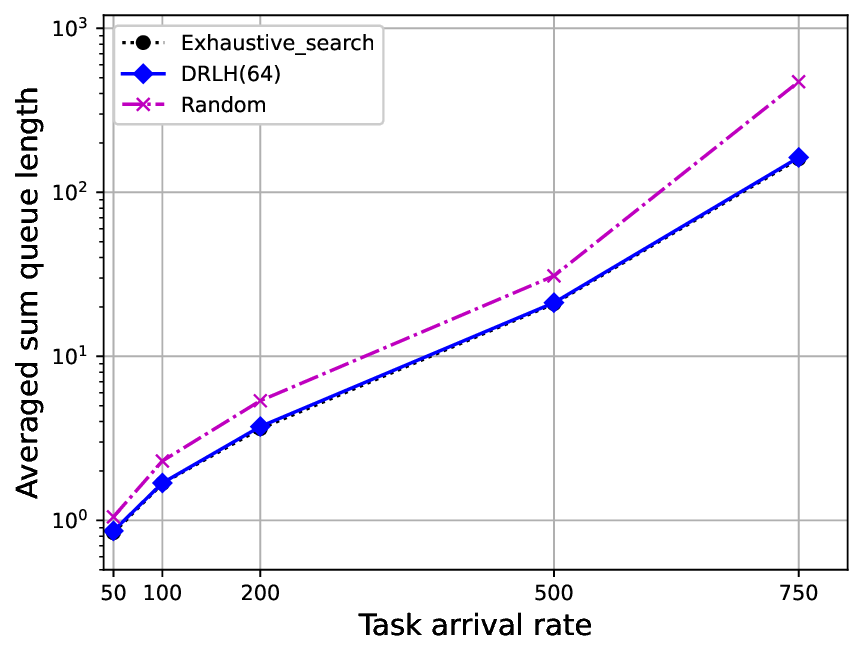}
}%
\centering
\subfigure[Queue length of the MEC server.]{
\includegraphics[width=0.66\columnwidth]{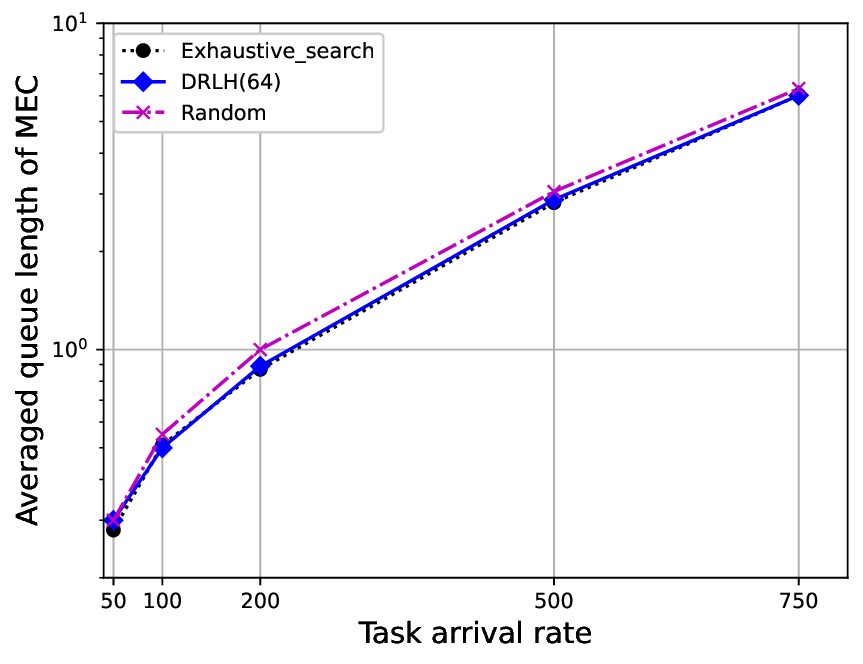}
}%
\centering
\subfigure[Weighted energy consumption of the system.]{
\includegraphics[width=0.66\columnwidth]{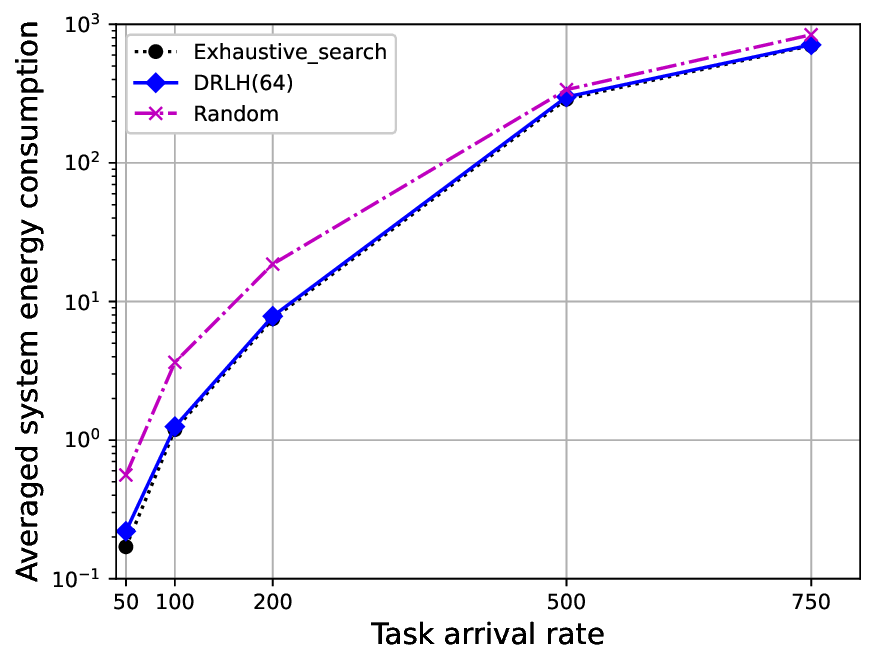}
}%
\caption{The averaged queue length and the energy consumption over the task arrival rate from 50 tasks/second to 750 tasks/second.}
\label{chp6:task_rate}
\end{figure*}



To verify the training performance of the proposed \textit{DRLH} framework, we compare the queue length, and the energy consumption of the following schemes and benchmarks. 
\begin{itemize}
\item {\textbf{Exhaustive search}: Instead of choosing the best action from the $N_a$ candidate actions, the exhaustive search scheme evaluates all offloading policies using the model-based optimization tool to find the optimal offloading policy and the computation capacity for users and the MEC server. According to the system settings, there are a total of 1960 offloading actions that satisfy the requirements.}
\item {\textbf{DRLH(64)}: The proposed \textit{DRLH} framework with $N_a = 64$ actions (approximate 3\% of the search space).}
\item {\textbf{DRLH(16)}: The proposed \textit{DRLH} framework with $N_a = 16$ actions (approximate 0.8\% of the search space).}
\item {\textbf{DRLH(8)}: The proposed \textit{DRLH} framework with $N_a = 8$ actions (approximate 0.4\% of the search space).}
\item {\textbf{Random}: The offloading policy is randomly chosen from the 1960 offloading actions.}
\end{itemize}

Meanwhile, to further evaluate the convergence performance and robustness of the proposed \textit{DRLH} framework in different situations, we consider the scenarios with different task arrival rates as follows.
\begin{itemize}
\item {\textbf{Scenario \uppercase\expandafter{\romannumeral1}}: In this scenario, we consider the reasonable task pressure with the mean arrival rate $\Lambda = 100$ tasks per second. The maximum queue length threshold for the end devices and the MEC server is the to $Q^{\rm{L}}_{\rm{max}} = 5$ and $Q^{\rm{E}}_{\rm{max}} = 1$, respectively. }
\item {\textbf{Scenario \uppercase\expandafter{\romannumeral2}}: In this scenario, we consider the high task pressure with the mean arrival rate $\Lambda = 750$ tasks per second. The maximum queue length threshold for the end devices and the MEC server is the to $Q^{\rm{L}}_{\rm{max}} = Q^{\rm{E}}_{\rm{max}} = \infty$.}
\end{itemize}

\begin{figure*}[htbp]
\centering
\subfigure[Queue length of users.]{
\includegraphics[width=0.66\columnwidth]{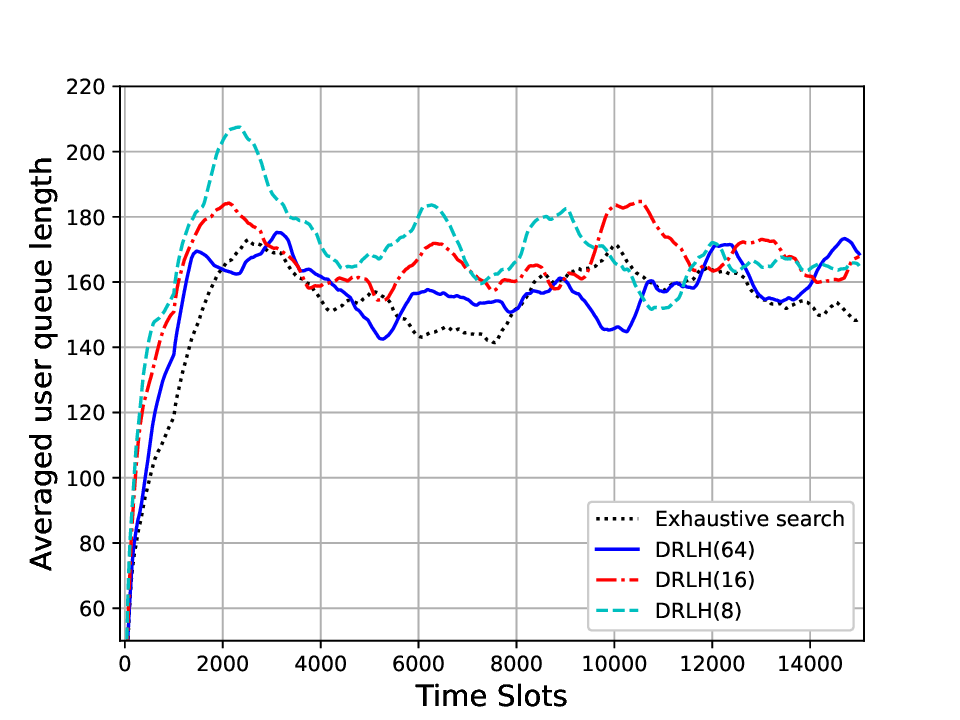}
}%
\centering
\subfigure[Queue length of the MEC server.]{
\includegraphics[width=0.66\columnwidth]{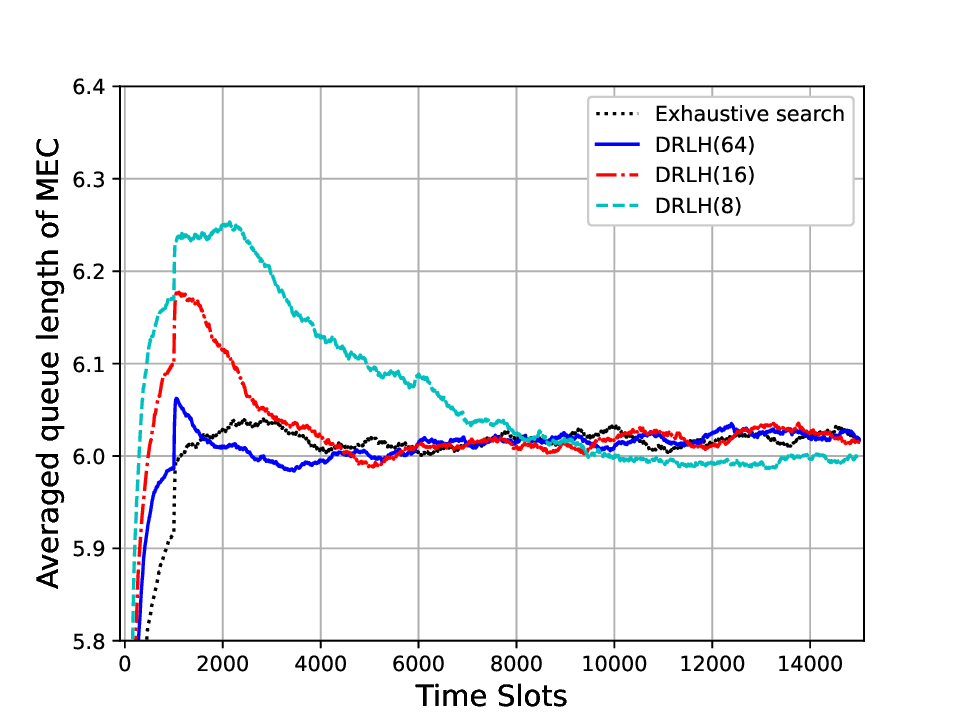}
}%
\centering
\subfigure[Weighted energy consumption of the system.]{
\includegraphics[width=0.66\columnwidth]{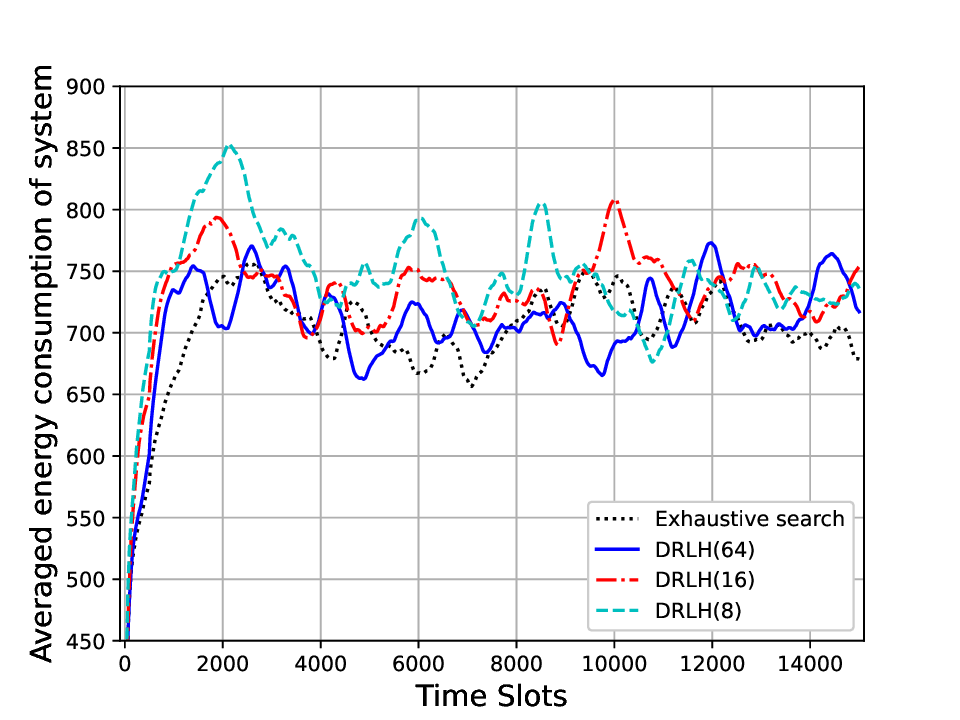}
}%
\caption{The averaged queue length and the energy consumption. The mean arrival rate $\Lambda$ of UEs is set to 750. The value of the curve is averaged every 1000 times slots to enhance the readability.}
\label{chp6:Amean_750}
\end{figure*}

\begin{figure}[htbp]
\centering
\subfigure[System sum queue length.]{
\includegraphics[width=0.8\columnwidth]{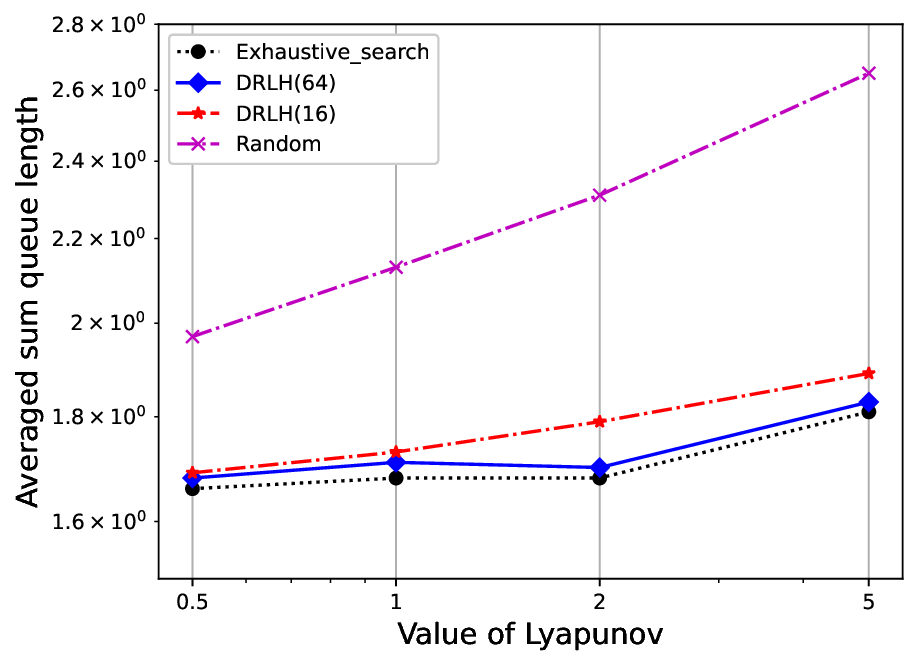}
}%
\centering

\subfigure[Weighted energy consumption of the system.]{
\includegraphics[width=0.8\columnwidth]{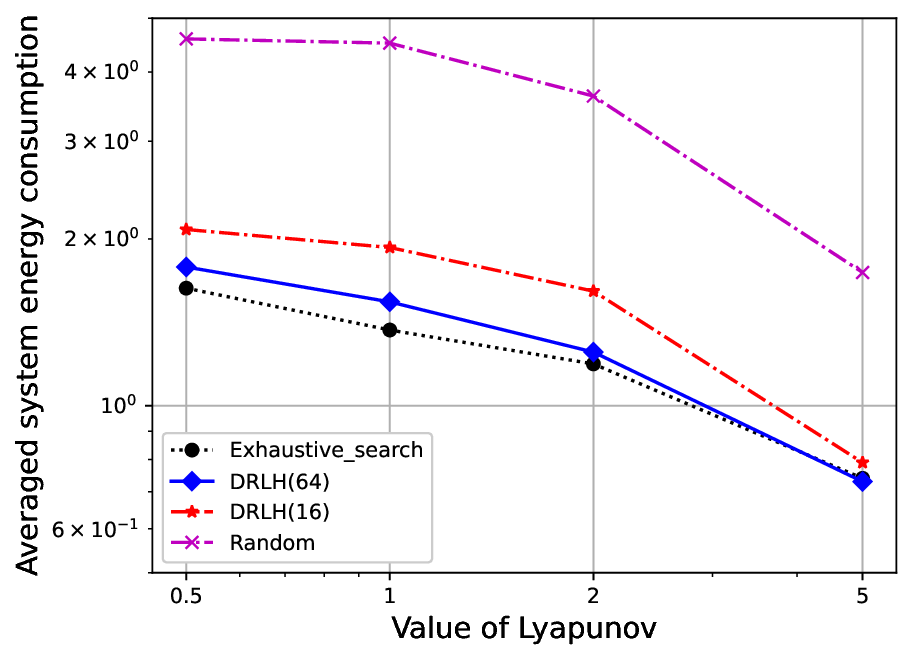}
}%
\caption{The averaged queue length and the energy consumption over the value of Lyapunov factors in \textbf{Scenario \uppercase\expandafter{\romannumeral1}}.}
\label{chp6:V_Amean_100}
\end{figure}

Fig.~\ref{chp6:loss} demonstrates the training performance of the proposed \textit{DRLH} frameworks. Particularly, we demonstrated training loss and testing loss to show the convergence performance of the proposed schemes. Particularly, the training loss represents the cross-entropy loss of the batch sampled from the replay memory, and the testing loss is the cross-entropy loss of the DNN output action and the optimal action selected by the model-based optimization algorithm.

As demonstrated by Fig.~\ref{chp6:loss}, the proposed \textit{DRLH} frameworks achieve consecutive decreases over the training time slots and remain stable when the training ends at 15000 time slots. Meanwhile, the frameworks can achieve convergence over different scenarios, showing the robustness of the training and scenario settings.

We first investigate the queue length and the system energy consumption with the task arrival rate, where the arrival rate is set to [50, 100, 200, 500, 750] tasks/second. As shown in Fig.~\ref{chp6:task_rate}, the queue length of users and the MEC servers grows nearly exponentially with the task arrival rate. The energy consumption of the cloud-edge-end network system, on the other hand, grows significantly when the arrival rate is low, while increasing less significantly when the task arrival rate is extreme. The maximum energy consumption of the system is bounded by the hardware and frequency constraints. Meanwhile, the proposed \textit{DRLH} scheme achieves almost the same performance as the exhaustive search benchmarks, verifying the effectiveness of the \textit{DRLH} scheme. 


We further verify the implementation performance and evaluate the degree of suboptimality of the \textit{DRLH} frameworks. It is noted that as the task arrivals follow the Poisson distribution, we focus on the average value of the queue length and the energy consumption metrics. Fig.~\ref{chp6:Amean_750} illustrates the performance of the proposed scheme in scenario \uppercase\expandafter{\romannumeral2}. {Note that the objective is to minimize the long-term energy consumption of the system, and the exhaustive search scheme only provides an optimal long-term performance benchmark, and we may observe a lower performance of exhaustive search in some time slots due to different state information.} Meanwhile, the DRLH(64) framework achieves almost the same performance as the exhaustive search algorithm, while reducing the computational complexity significantly (only 3\% actions are selected and evaluated). Meanwhile, the performance of the DRLH(16) and the DRLH(8) schemes show the descending order, due to the reduced searching space and the computational complexity. It is noted that even the DRLH(8) scheme meets the queue length constraint for the end devices and the MEC server, further verifying the robustness of the proposed framework.

Particularly, guided by the critic module, the \textit{DRLH} schemes show a learning manner to achieve the stability of the queues. Particularly, in Fig.~\ref{chp6:Amean_750} (a) and (b), we can observe that the queue length increases significantly at the beginning of the training stage. Due to the unawareness of the model of the system and a large number of arrival tasks, the queue length of the \textit{DRLH} schemes cannot remain stable at the beginning of the training stage. On the other hand, the exhaustive search scheme maintains the stability of the queue during the whole training stage. 

The learning ability of the DNN enables the model-free NN to adapt to the scenario and find near-optimal solutions, which is further demonstrated in Fig.~\ref{chp6:Amean_750} (a) and (b). Although the queue length increases significantly at the beginning, the queues of \textit{DRLH} frameworks decrease with the process of the training over time slots. Specifically, the queue length of the DRLH(64) scheme for the MEC server decreases and remains at a stable value over 2000 time slots. The DRLH(16) and DRLH(8) schemes, on the other hand, achieve queue stability over 4000 and 6000 time slots, respectively. The reason is that the DRLH(64) explores a larger action space, and can find the near-optimal solutions more quickly than the DRLH(16) and DRLH(8) schemes. 

\begin{figure}[tp]
\centering
\subfigure[System sum queue length.]{
\includegraphics[width=0.8\columnwidth]{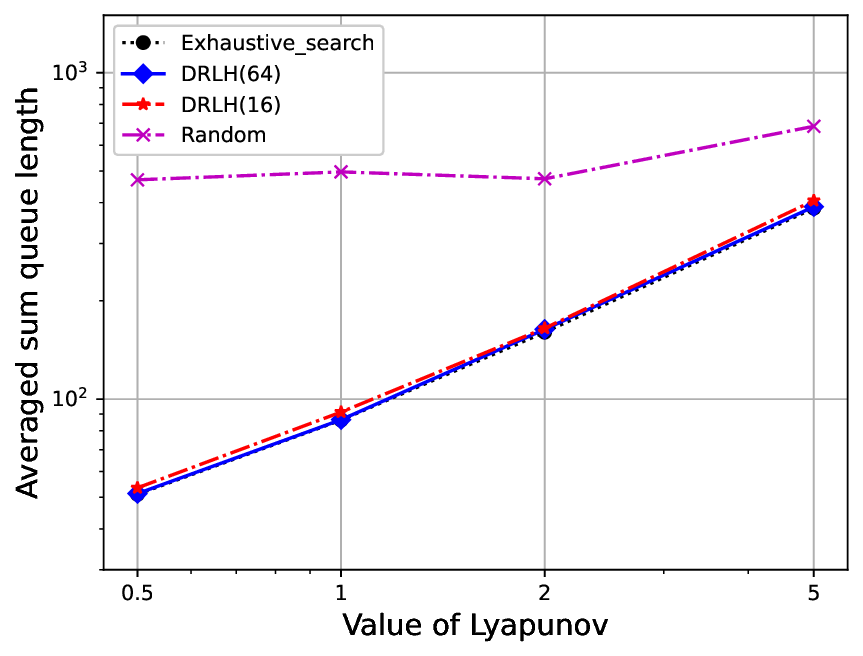}
}%

\centering
\subfigure[Weighted energy consumption of the system.]{
\includegraphics[width=0.8\columnwidth]{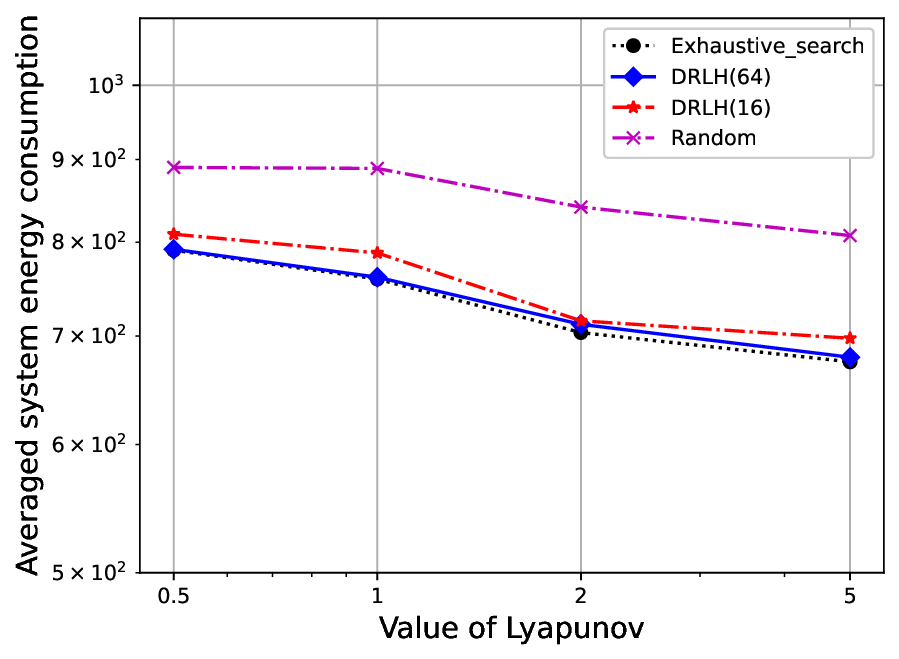}
}%
\caption{The averaged queue length and the energy consumption over the value of Lyapunov factors in \textbf{Scenario \uppercase\expandafter{\romannumeral2}}.}
\label{chp6:V_Amean_750}
\end{figure}

In Fig.~\ref{chp6:V_Amean_100} and Fig.~\ref{chp6:V_Amean_750}, we further investigate the influence of the Lyapunov factor $v$, which represents the importance of the power consumption versus the queue length. The queue length over the Lyapunov factor is demonstrated in Fig.~\ref{chp6:V_Amean_100} (a) and (b) for scenario \uppercase\expandafter{\romannumeral1}, and in Fig.~\ref{chp6:V_Amean_750} (a) and (b) for scenario \uppercase\expandafter{\romannumeral2}, respectively. It is observed that the proposed \textit{DRLH} scheme outperforms the random scheme in terms of the queue length and the energy consumption in both scenarios, verifying the effectiveness and robustness of the \textit{DRLH} framework. Meanwhile, the overall queue length increases with the larger value of $v$. However, the energy consumption of the scenario \uppercase\expandafter{\romannumeral1} remains stable over different values of $v$, which reveals that the energy consumption is hard to reduce in \uppercase\expandafter{\romannumeral1}, and in such scenario a smaller $v$ should be choice for reducing the queue backlog and task execution delay. 

On the other hand, in Fig.~\ref{chp6:V_Amean_750} for scenario \uppercase\expandafter{\romannumeral2}, it is observed that the power consumption of the system can be reduced effectively with the larger $v$, while the average queue length increasing significantly. In such scenarios, the value of $v$ should refer to the QoS and QoE preference of end devices, so as to achieve the tradeoff between the queue length (task execution delay) and the energy consumption. 

\begin{figure}[t]
\centering
{
\includegraphics[width=0.8\columnwidth]{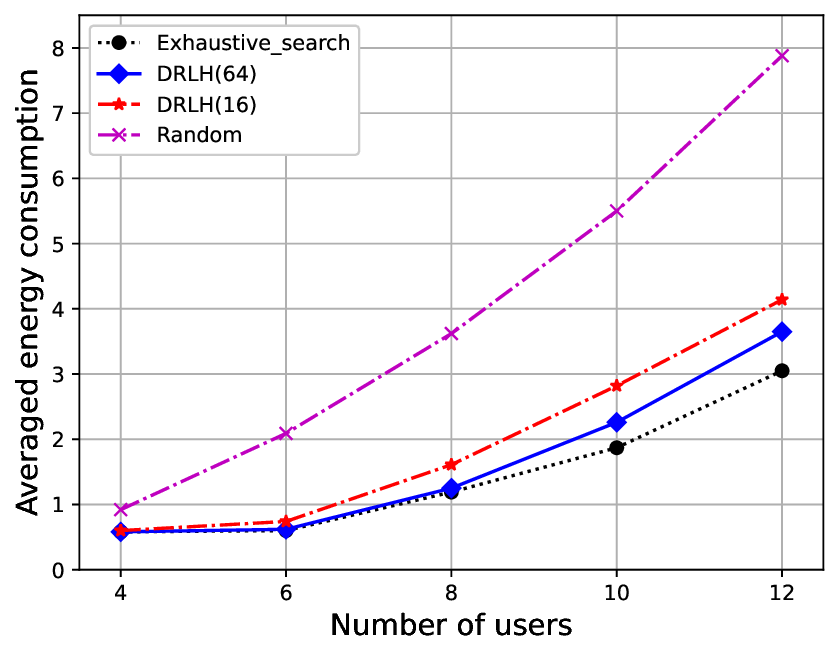}
}%
\caption{Weighted energy consumption of the system in \textbf{Scenario I}.}
\label{chp6:num_users}
\end{figure}

{To evaluate the scalability of the proposed framework, we further implement \textit{DRLH} scheme over the different numbers of end users. Intuitively in Fig.~\ref{chp6:num_users}, the average energy consumption of the system increases with more users, as the number of income tasks increases with users. The search space for 4, 6, 8, 10, and 12 users increase significantly, from 6, 225, 1960, 9450, to 32670, respectively. The computational complexity of the exhaustive search scheme is enormous and cannot be adapted to next-generation networks with massive connectivity. On the other hand, the proposed \textit{DRLH} framework, only needs to evaluate a fixed number of actions, and achieves near-optimal performance with much lower computational complexity. It is noted that with the searching space increases, the actions evaluated by the proposed \textit{DRLH} framework make up only a small proportion of all possible actions, and the gap between the exhaustive search and the \textit{DRLH} framework may increase. In different scenarios, we should implement an appropriate value of the number of candidate actions $N_a$ to strike a performance-complexity tradeoff.}

\section{Conclusion}
\label{chp6:conclusion}

In this paper, a cloud-edge-end collaborative network has been proposed to execute intelligent tasks through the collaborative computing network. Various computational offloading designs were made to enhance the computational performance in the heterogeneous system. To realize a long-term optimization of the energy consumption of the system, a Lyapunov-guided methodology was employed to convert the multi-stage problem into deterministic problems that are resolvable within individual time slots. A hybrid framework that integrated the model-free deep reinforcement learning algorithm and model-based optimization algorithm was proposed to optimize the communication resources and computation resources jointly. Numerical results demonstrated that the proposed resource allocation framework achieved near-optimal long-term energy efficiency while maintaining the stability of task queues and different numbers of users, {and the tradeoff between energy and latency, performance and complexity should be made in practical implementations.} In conclusion, by offloading the local computations to the edge and cloud servers and leveraging the task-oriented semantic communication, the computational and communication resources can be coordinated and flow across the heterogeneous network entities, enabling the implementation of the emerging artificial intelligent applications and the massive connectivity of modern communication systems.

\appendices
\section{Proof of Theorem~\ref{chp6:theorem_1}}
\label{appendix_1}
We define that the $\boldsymbol{Q}^{\rm{L}} = \{Q^{\rm{L}}_i\}_{i \in I}$, $\boldsymbol{Q}^{\rm{E}} = \{Q^{\rm{E}}_i\}_{i \in I}$, $\boldsymbol{Z}^{\rm{L}} = \{Z^{\rm{L}}_i\}_{i \in I}$, $\boldsymbol{Z}^{\rm{E}} = \{Z^{\rm{E}}_i\}_{i \in I}$ are the system state variables. The Lyapunov function ${\cal L}(\cdot)$ and drift function $\Delta(\cdot)$ can be denoted as a similar way as~(\ref{chp6:Lyapunov_function}) and~(\ref{chp6:Lyapunov_drift}), respectively. 

The upper bound of the drift for each of the above queue length related variables is proved and given by the following lemmas. First we denote that $[x]^{+} = \max \{x, 0\}$, and $\mu^{\rm{L}}_{\max}$, $u^{\rm{L}}_{\max}$, $u^{\rm{E}}_{\max}$, $u^{\rm{C}}_{\max}$, $\Lambda_{\max}$ represent the upper bound of the task execution rate $\mu^{\rm{L}}$, the local computing rate $u^{\rm{L}}$, the MEC offloading rate $u^{\rm{E}}$, the MCC offloading rate $u^{\rm{C}}$, and the task arrival rate $\Lambda$, respectively.

\begin{lemma}
\label{lemma:QL}
The drift function for $\boldsymbol{Q}^{\rm{L}}(t)$ is bounded as 
\begin{equation}
\Delta(\boldsymbol{Q}^{\rm{L}}(t)) \leq B_1 - \sum_{i \in I} \mathbb {E}\left[Q^{\rm{L}}_i(t)(\mu^{\rm{L}}_i(t)-\Lambda_i(t))|\boldsymbol{\Theta(t)}\right],
\label{lemma_QL}
\end{equation}
where $B_1 = \frac{1}{2} \sum_{i \in I} [(\mu^{\rm{L}}_{\max})^2 + (\Lambda_{\max})^2]$.

\begin{proof} 
According to the definition of the Lyapunov drift function, the $\Delta(Q^{\rm{L}}_i(t)) = \frac{1}{2} \left[Q^{\rm{L}}_i(t+1)^2-Q^{\rm{L}}_i(t)^2\right]$. Meanwhile, the $Q^{\rm{L}}_i(t+1)^2 = \left[\left(Q^{\rm{L}}_i(t)-\mu^{\rm{L}}_i(t)\right)^{+} + \Lambda_i\right]^2 = \left[Q^{\rm{L}}_i(t)-\mu^{\rm{L}}_i(t) + \Lambda_i\right]^2 = (Q^{\rm{L}}_i(t))^2 - 2Q^{\rm{L}}_i(t)(\mu^{\rm{L}}_i(t) - \Lambda_i(t)) + (\mu^{\rm{L}}_i(t) - \Lambda_i(t))^2$.

Hence, the upper bound of the Lyapunov drift function for $Q^{\rm{L}}_i(t)$ can be expressed by

\begin{equation}
\begin{split}
&\frac{1}{2} \left((Q^{\rm{L}}_i(t+1))^2 - (Q^{\rm{L}}_i(t))^2\right) \\ &= - Q^{\rm{L}}_i(t)(\mu^{\rm{L}}_i(t) - \Lambda_i(t)) +\frac{1}{2} (\mu^{\rm{L}}_i(t) - \Lambda_i(t))^2 \\
&+\leq - Q^{\rm{L}}_i(t)(\mu^{\rm{L}}_i(t) - \Lambda_i(t)) +\frac{1}{2} \left[(\mu^{\rm{L}}_{\max})^2- (\Lambda_{\max})^2\right]. 
\end{split}
\end{equation}

By computing the conditional expectation on both sides of the inequality and aggregating the results across all devices $i \in I$, we obtain (\ref{lemma_QL}).
\end{proof}
\end{lemma}

\begin{lemma}
\label{lemma:ZL}
The Lyapunov drift function for $\boldsymbol{Z}^{\rm{L}}(t)$ is bounded as 
\begin{equation}
\begin{aligned}
\Delta(\boldsymbol{Z}^{\rm{L}}(t)) & \leq B_2- \sum_{i \in I} \mathbb {E}\\ &\left[Z^{\rm{L}}_i(t)(\mu^{\rm{L}}_i(t)-\Lambda_i(t)+Q^{\rm{L}}_{\max}-Q^{\rm{L}}_i(t))|\boldsymbol{\Theta(t)}\right],
\label{lemma_ZL}
\end{aligned}
\end{equation}
where 
\begin{equation}
\begin{aligned}
B_2 &= \frac{1}{2} \sum_{i \in I} \left[(\mu^{\rm{L}}_{\max})^2 + (\Lambda_i(t))^2+ (Q^{\rm{L}}_i(t))^2 + (Q^{\rm{L}}_{\max})^2\right] \\ &+\mu^{\rm{L}}_{\max}Q^{\rm{L}}_{\max} + \Lambda_i(t)Q^{\rm{L}}_i(t).
\end{aligned}
\end{equation}

\begin{proof} 
According to the definition of $Z^{\rm{L}}_i(t+1)$ in~(\ref{chp6:v_queue}), we have 
\begin{equation}
\begin{split}
    Z^{\rm{L}}_i(t+1)^2 \leq &\left(Z^{\rm{L}}_i(t) + Q^{\rm{L}}_i(t+1) - Q^{\rm{L}}_{\max}\right)^2 \\ = 
    &\left(Z^{\rm{L}}_i(t) + Q^{\rm{L}}_i(t) - \mu^{\rm{L}}_i(t) + \Lambda_i(t)- Q^{\rm{L}}_{\max}\right)^2 \\ 
    = &(Z^{\rm
    {L}}_i(t))^2 - 2Z^{\rm{L}}_i(t)(\mu^{\rm{L}}_i(t) -  \Lambda_i(t) - Q^{\rm{L}}_i(t) + Q^{\rm{L}}_{\max}) \\ &+ (\mu^{\rm{L}}_i(t) -  \Lambda_i(t) - Q^{\rm{L}}_i(t) + Q^{\rm{L}}_{\max})^2.
\end{split}
\end{equation}
Thus the upper bound of the Lyapunov drift function for $Q^{\rm{L}}_i(t)$ can be expressed by 

\begin{equation}
\begin{split}
&\frac{1}{2} \left((Z^{\rm{L}}_i(t+1))^2 - (Z^{\rm{L}}_i(t))^2\right) \\ =  &Z^{\rm{L}}_i(t)(\mu^{\rm{L}}_i(t) -  \Lambda_i(t) - Q^{\rm{L}}_i(t) + Q^{\rm{L}}_{\max}) \\
& + \frac{1}{2} \left[\left(\mu^{\rm{L}}_i(t) + Q^{\rm{L}}_{\max}\right) -  \left(\Lambda_i(t)  + Q^{\rm{L}}_i(t)\right) \right]^2
\\ \leq  &Z^{\rm{L}}_i(t)(\mu^{\rm{L}}_i(t) -  \Lambda_i(t) - Q^{\rm{L}}_i(t) + Q^{\rm{L}}_{\max}) \\
& + \frac{1}{2} \left[(\mu^{\rm{L}}_i(t))^2 + (Q^{\rm{L}}_{\max})^2 + (\Lambda_i(t))^2  + (Q^{\rm{L}}_i(t))^2 \right] \\ &+\mu^{\rm{L}}_{\max}Q^{\rm{L}}_{\max} + \Lambda_i(t)Q^{\rm{L}}_i(t).
\end{split}
\end{equation}

By computing the conditional expectation on both sides of the inequality and aggregating the results across all devices $i \in I$, we obtain (\ref{lemma_ZL}).
\end{proof}
\end{lemma}

\begin{lemma}
\label{lemma:QL2}
The drift function for $\boldsymbol{Q}^{\rm{E}}(t)$ is bounded as 
\begin{equation}
\Delta(\boldsymbol{Q}^{\rm{E}}(t)) \leq B_3 - \sum_{i \in I} \mathbb {E}\left[Q^{\rm{E}}_i(t)(\mu^{\rm{E}}_i(t)-u^{\rm{E}}_i(t))|\boldsymbol{\Theta(t)}\right],
\label{lemma_QLE}
\end{equation}
where $B_3 = \frac{1}{2} \sum_{i \in I} [(\mu^{\rm{E}}_{\max})^2 + (u^{\rm{E}}_{\max})^2]$.

\begin{proof} 
The proof is similar to that of \textbf{Lemma~\ref{lemma:QL}}.
\end{proof}
\end{lemma}

\begin{lemma}
\label{lemma:ZLE}
The drift function for $\boldsymbol{Z}^{\rm{E}}(t)$ is bounded as 
\begin{equation}
\begin{aligned}
    \Delta(\boldsymbol{Z}^{\rm{E}}(t)) \leq & B_4 - \sum_{i \in I} \mathbb {E} \\
    & \left[Z^{\rm{E}}_i(t)(\mu^{\rm{E}}_i(t)-u^{\rm{E}}_i(t)+Q^{\rm{E}}_{\max}-Q^{\rm{E}}_i(t))|\boldsymbol{\Theta(t)}\right],
\end{aligned}
\end{equation}
where
\begin{equation}
\begin{aligned}
B_4 &= \frac{1}{2} \sum_{i \in I} \left[(\mu^{\rm{E}}_{\max})^2 + (u^{\rm{E}}_i(t)))^2+ (Q^{\rm{E}}_i(t))^2 + (Q^{\rm{E}}_{\max})^2\right] \\ &+\mu^{\rm{E}}_{\max}Q^{\rm{E}}_{\max} + u^{\rm{E}}_{\max}Q^{\rm{E}}_i(t).
\end{aligned}
\end{equation}

\begin{proof} 
The proof is similar to that of \textbf{Lemma~\ref{lemma:ZL}}.
\end{proof}
\end{lemma}

According to the \textbf{Lemma~\ref{lemma:QL}} to \textbf{Lemma~\ref{lemma:ZLE}}, the overall Lyapunov drift function can be thereby denoted as $\Delta(\boldsymbol{\Theta(t)}) = \Delta(\boldsymbol{Q}^{\rm{L}}(t)) + \Delta(\boldsymbol{Q}^{\rm{E}}(t)) + \Delta(\boldsymbol{Z}^{\rm{L}}(t)) + \Delta(\boldsymbol{Z}^{\rm{E}}(t))$. By summing up the results, we obtain the drift-plus-penalty as~\ref{chp6:Lyapunov_drift_penalty}, where $\hat{B} = B_1 + B_2 + B_3 + B_4 + \sum_{i \in I} \left[Z^{\rm{L}}_i(t)\left(Q^{\rm{L}}_i(t)-Q^{\rm{L}}_{\max}\right) + Z^{\rm{E}}_i(t)\left(Q^{\rm{E}}_i(t)-Q^{\rm{E}}_{\max}\right)\right]$. As $\hat{B}$ only consists of constant terms from the observation, we can put it aside from the optimization of the variables.
\bibliography{Reference}

\clearpage
\end{document}